\documentclass[11pt,english,table]{article}
\usepackage{mathptmx}

\usepackage[T1]{fontenc}
\usepackage[latin9]{inputenc}
\usepackage{geometry}
\usepackage{float}
\usepackage{bm}
\usepackage{amsmath}
\usepackage{amsthm}
\usepackage{amssymb}
\usepackage{graphicx}
\usepackage{rotfloat}
\usepackage{setspace}
\usepackage[authoryear]{natbib}
\makeatletter
\theoremstyle{plain}
\newtheorem{thm}{\protect\theoremname}
\theoremstyle{plain}
\newtheorem{cor}{\protect\corollaryname}

\setcitestyle{round}
\usepackage{lscape}
\usepackage{longtable}
\usepackage{xcolor}
\usepackage{rotating}

\usepackage{array}
\usepackage{tabularx}\usepackage{multirow}\usepackage{booktabs}
\usepackage{ragged2e}\newcolumntype{C}[1]{>{\centering\arraybackslash}p{#1}}
\usepackage{rotfloat}
\newcolumntype{J}[1]{>{\justify\arraybackslash}p{#1}}
\newcolumntype{R}[1]{>{\RaggedLeft\arraybackslash}p{#1}}
\newcolumntype{Q}[1]{>{\columncolor{Gray}\RaggedLeft\arraybackslash}p{#1}}
\newcolumntype{L}[1]{>{\RaggedRight\arraybackslash}p{#1}}
\newcolumntype{G}{@{\extracolsep{0.5cm}}l@{\extracolsep{0pt}}}%
\newcolumntype{P}[1]{>{\centering\arraybackslash}p{#1}}
\newcolumntype{Y}{>{\centering\arraybackslash}X}
\newcommand{\nhphantom}[1]{\sbox0{#1}\hspace{-\the\wd0}} 

\AtBeginDocument{

}

\usepackage[colorlinks,
            linkcolor=blue,
            anchorcolor=blue,
            citecolor=blue]{hyperref}

\usepackage{babel}

\makeatother

\usepackage{babel}
\providecommand{\corollaryname}{Corollary}
\providecommand{\theoremname}{Theorem}

\begin{document}
\title{Option Pricing with State-dependent Pricing Kernel\emph{\normalsize{}\medskip{}
}}
\author{\textbf{Chen Tong}$^{a}$\textbf{$\quad$$\quad$Peter Reinhard Hansen}$^{b}$\thanks{Correspondence author: Peter Reinhard Hansen (hansen@unc.edu). We would like to acknowledge the helpful comments and suggestions from Bart Frijns (the editor), Robert I. Webb, Biao Guo (discussant), and the audience at the 2021 International Conference on Derivatives and Capital Markets. We thank Emily Dyckman for proofreading the manuscript. Zhuo Huang acknowledges the financial support from the National Natural Science
Foundation of China (71671004). Chen Tong's research was supported by the Fundamental Research Funds for the Central Universities (20720221025).}\textbf{$\quad$$\quad$Zhuo Huang}$^{c}$\bigskip{}
\\
 {\normalsize{}$^{a}$}\emph{\normalsize{}Department of Finance, School
of Economics \& Wang Yanan Institute} \emph{\normalsize{}for}\\
\emph{\normalsize{} Studies in Economics (WISE), Xiamen University}\\
 {\normalsize{}$^{b}$}\emph{\normalsize{}University of North Carolina
\& Copenhagen Business School}\\
 {\normalsize{}$^{c}$}\emph{\normalsize{}National School of Development,
Peking University}\\
\emph{\normalsize{}\medskip{}
 }}
\date{\emph{\normalsize{}\today}}
\maketitle
\begin{abstract}
We introduce a new volatility model for option pricing that combines
Markov switching with the Realized GARCH framework. This leads to a 
novel pricing kernel with a state-dependent variance risk premium and a pricing
formula for European options, which is derived with an analytical approximation method.
We apply the Markov switching Realized GARCH
model to S\&P 500 index options from 1990 to 2019 and find that investors'
aversion to volatility-specific risk is time-varying. The proposed
framework outperforms competing models and reduces (in-sample and out-of-sample) option pricing
errors by 15\% or more.
\bigskip{}
\end{abstract}
\textit{\small{}{\noindent}Keywords:}{\small{} Option Pricing, Realized
GARCH, Regime-switching, Variance Risk Premium, Edgeworth Expansion}{\small\par}

\noindent \textit{\small{}{\noindent}JEL Classification:}{\small{}
G12, G13, C51, C52}{\small\par}

\bigskip{}

\section{Introduction }

Risk aversion is a fundamental concept in economic theory with uncertainty.
It is embedded in the pricing kernel that bridges the physical
probability measure, $\mathbb{P}$, with the risk-neutral measures,
$\mathbb{Q}$.\footnote{The risk-neutral measure (or equivalent martingale measure) is a probability measure such that each share price is exactly equal to the risk-free discounted expectation of the share price under this measure.} 
In the classical asset pricing model with a risk-averse investor,
see \citet{Lucas1978}, the pricing kernel is monotonically declining
in aggregate wealth. In practice, the pricing kernel is unknown but
can be estimated using a variety of econometric methods, see e.g.
\citet{AitSahaliaLo2000}, \citet{Jackwerth2000} and \citet{RosenbergEngle2002}.
When an empirical pricing kernel is plotted against the market return,
it often has a upward sloping region. This is contrary
to the classical model and has become known as a \textit{pricing kernel
puzzle}. The empirical pricing kernel is typically found to be U-shaped when
estimated over a long sample period, see e.g. \citet{BakshiMadanPanayotov2010}
and \citet{ChristoffersenHestonJacobs2013}.

A popular explanation for the \textit{\emph{pricing kernel puzzle}}
is the existence of additional risk factors beyond the equity risk
premium. Compensation for additional risk factors can influence the
projection onto the return dimension and induce the pricing kernel
puzzle. A possible risk factor, which is supported by the empirical evidence,
is the \emph{variance risk premium}. The squared VIX is a measure
of the squared variation under $\mathbb{Q}$ and it is, on average,
larger than empirical measures of return variance under $\mathbb{P}$.
When volatility is larger under the risk-neutral measure than under
the physical measure it implies a negative variance risk premium,
which can explain the observed U-shaped in empirical pricing kernels.
An important contribution to this literature was made in \citet{ChristoffersenHestonJacobs2013},
who proposed a variance-dependent pricing kernel with a variance risk
premium in addition to the equity risk premium. 

While pricing kernels are typically found to be U-shaped, there is
also evidence that their shapes are unstable over time. In fact, there
are periods when the U-shaped pricing kernel disappeared, and exhibited hump-shaped or even inverted U-shaped. Examples of this can be
seen in \citet{ChristoffersenHestonJacobs2013} for their semi-parametric
estimates during the year 2004 to 2007. For the same sample period, 
\citet{GrithHardleKratschmer2017} estimated a hump-shaped pricing
kernel with DAX 30 index options. A related approach is to estimate
the probability weighting function, as in \citet{PolkovnichenkoZhao2013}.
They estimate the probability weighting function to have a regular
S-shaped in the same period (2004-2007) using S\&P 500 index options, as opposed to an inverse S-shaped, which they estimate for most periods.
The latter corresponds to a U-shaped pricing kernel.\footnote{An inverse S-shaped implies that investors tend to overweight low-probability
events while they underweight the likelihood of events with high probability.
The opposite is true with a regular S-shape.} Similarly, \citet{KieselRahe2017} found a slight positive variance
risk premium from mid-2004 to mid-2007. So, these studies suggest
that the variance risk aversion was relatively low during this period.

In classical asset pricing models, the risk aversion parameter is
constant over time, and this greatly simplifies the implementation
of these models. If risk aversion is time-varying, but incorrectly
assumed to be constant, then this can induce large pricing errors
because the misspecified model leads to an incorrect option pricing
formula. The assumption of constant risk aversion is contradicted
by empirical evidence, and several theoretical models with time-varying
risk aversion have been proposed in the literature, see e.g. \citet{CampbellCochrane1999},
\citet{Li2007}, \citet{Chabi-YoGarciaRenault2008},
and \citet{BekertEngstromXu2019}. 

In this paper, we propose a novel state-dependent pricing kernel with
a variance risk premium and time-varying risk aversion. First, we
introduce a new discrete-time volatility model within the Realized
GARCH framework of \citet{HansenHuangShek:2012}, in which a latent
state follows a \textit{hidden} Markov switching process. The same
Markov switching process is used to introduce time-variation in the
pricing kernel, and this leads to a pricing kernel with a state-dependent
variance risk premium. The framework has a great deal of flexibility
in terms of the statistical model for the observed variables under
$\mathbb{P}$ and in terms of the plasticity of the pricing kernel.
It is, nevertheless, possible to derive the corresponding pricing
formula for European options by means of an analytical approximation
method. The Markov-switching Realized GARCH model is relatively simple
to estimate by quasi maximum likelihood, and the latent states can be inferred from
the observed realized volatility measures and returns. The Realized GARCH framework is well suited for this problem
because a key component in the model is a volatility-specific shock.
This shock is inferred from the difference between the realized volatility
measure and its conditional expectation, and it can be incorporated
in the pricing kernel to include a variance risk premium.\footnote{A growing literature is exploring ways to utilize realized volatility measures for derivatives pricing, see e.g. \citet{CorsiFusariVecchia2013,ChristoffersenFeunouJacobsMeddahi2014,MajewskiBormettiCorsi2015,HuangWangHansen2017,HuangTongWang2019,TongHuang2021}.} Estimation that includes the estimation of the parameters in the pricing
kernel is more involved because it relies on the observed option prices.\footnote{\citet{HansenTong2021} also introduces an option
pricing model with time-varying volatility risk aversion. Their framework
is different in two important ways. First, they build on the Heston-Nandi
GARCH model (\citealp{HestonNandi2000}). Second, they use a score-driven
model, see \citet{Creal2013}, to model time-variation in risk aversion.}

In this paper, we conduct an extensive empirical analysis with a large
panel of S\&P 500 index option prices based on 30 years of data (from
1990 to 2019). We find the volatility risk aversion to be time-varying.
Investors tend to be more risk-averse during periods with high volatility,
but there are also periods where investors have a slight appetite for
the variance risk, such as during the low-volatility periods: 1993-1995,
2004-2007, and 2014-2017. This is consistent with other studies that
report a positive volatility risk premium during these periods. On
average, we find the volatility risk premium to be negative, which
is consistent with existing literature. In terms of option pricing
performance, our empirical results show that the proposed framework
outperforms other benchmarks by reducing option pricing errors by
15\% or more. These reductions in pricing errors are found both in-sample
and out-of-sample.

The successful fusion of a GARCH model with a Markov switching
structure is an accomplishment that deserves some commenting. Estimation
of GARCH models with Markov switching is typically marred with complications.
For instance, the path-dependence problem makes it nearly impossible
to evaluate the sample likelihood function. The solution by \citet{Cai1994}
and \citet{HamiltonSusmel1994} is only valid within an ARCH specification.
Other approaches, such as those by \citet{Gray1996}, and \citet{Klaassen2002},
are not suitable for asset pricing due to the difficulties in risk
neutralization.\footnote{Most option pricing models under the specification
of \citet{Gray1996} are constructed without specifying risk premium, see e.g. \citet{SatoyoshiMitsui2011} and \citet{DaoukGuo2004}.}
\citet{ChenHung2010} directly assume a Markov-switching GARCH process for
returns under the risk-neutral measure and obtain option prices by
a lattice method, volatility discretization, and Monte Carlo simulation.
However, their model is not amenable to estimation and the authors
resort to calibration instead. \citet{ElliottSiuChan2006} develop
a method for a Heston-Nandi GARCH model with Markov switching, but
it requires the Markov states to be observable. For this reason, there
is not much empirical literature on option pricing based
on a GARCH model with a hidden Markov switching structure.\footnote{There are several Markov switching models for option pricing without
GARCH structures, see e.g. \citet{Duan2002}, \citet{DonaldDasMotwani2006},
\citet{LiewSiu2010}, and \citet{ShenFanSiu2014}.} 

The key to our successful coupling of a hidden Markov switching model
with a GARCH-type model is the presence of the realized measures of
volatility in the model. By providing accurate information about the contemporaneous
volatility level, the realized measure adds valuable information about
the latent state, and the Realized GARCH model maintains the structure
that permits straightforward estimation based on returns and realized
measures. The most complicated part of the framework relates to the
analytical approximation for option pricing, because it requires many
terms to be computed. The expressions for these terms are computed
in the Appendix and are plugged into the generic expressions derived
in \citet[p. 104]{DuanGauthierSimonato1999}.

The rest of the paper is organized as follows. In Section 2, we propose
the theoretical model under the physical measure, $\mathbb{P}$, the
risk neutralization process under a state-dependent pricing kernel,
and details about the estimation of the Markov switching Realized GARCH
model. Section 3 introduces the analytical approximation formula for
European call options. Section 4 presents the competing models used
in our empirical comparisons. Section 5 describes the joint estimation
method. All empirical results are presented in Section 6 including
the summary statistics, parameters estimation, and in-sample and out-of-sample
option pricing performance. All relevant proofs are presented in 
Appendix A, and Appendix B derives the terms needed for option pricing.

\section{The Model\label{sec:Model}}

Let $\mathcal{F}_{t}=\ensuremath{\sigma}(\{R_{\tau},x_{\tau}\},\tau\leq t)$
denote the natural filtration, where $R_{t+1}=\log\left(S_{t+1}/S_{t}\right)$
is the daily log-return, and $x_{t}$ is the realized measure of volatility.
The latter is, in our empirical analysis, computed from high-frequency
data with the Realized Kernel by \citet{BNHLS:2008}.

\subsection{The Realized GARCH Model ($\mathbb{P}$), }

The Realized GARCH framework was introduced by \citet{HansenHuangShek:2012}.
In this paper, we build on the variant proposed in \citet{HansenHuang:2016}
that was also used for option pricing  in \citet{HuangWangHansen2017}.
The dynamic properties of returns, the conditional variance, $h_{t+1}=\mathrm{var}(R_{t+1}|\mathcal{F}_{t})$,
and the realized volatility measure are given by the equations:
\begin{eqnarray}
R_{t+1} & = & r+\lambda\sqrt{h_{t+1}}-\tfrac{1}{2}h_{t+1}+\sqrt{h_{t+1}}z_{t+1},\label{eq:return}\\
\log h_{t+1} & = & \omega+\beta\log h_{t}+\gamma\log x_{t}+\tau_{1}z_{t}+\tau_{2}(z_{t}^{2}-1),\label{eq:garch}\\
\log x_{t} & = & \xi+\phi\log h_{t}+\delta_{1}z_{t}+\delta_{2}(z_{t}^{2}-1)+\sigma u_{t}.\label{eq:measurement}
\end{eqnarray}
The intercept in the return equation, $r$, denotes the risk-free
rate and $\lambda$ is the equity risk premium. The stochastic properties
are driven by two i.i.d standard normally distributed ``innovations'',
$z_{t}$ and $u_{t}$, that represent return and volatility shocks,
respectively.

Realized GARCH models are characterized by the measurement equation,
(\ref{eq:measurement}), that defines the relationship between the
conditional variance, $h_{t}$, and the realized measure $x_{t}$.
Unlike the conventional GARCH model, this model has two distinct innovations,
$z_{t}$ and $u_{t}$, where the second is similar to the random innovation
for volatility in stochastic volatility models. However, because the
Realized GARCH model is an observation-driven model, it is simpler
to estimate than stochastic volatility models. For the purpose of
derivative pricing, the Realized GARCH structure is especially valuable
because the two innovation terms can be used as risk factors, as we
explore below.

\subsection{The Markov-switching Realized GARCH Model ($\mathbb{P}$)}

Here, we introduce a new volatility model based on the Realized GARCH
model above. Specifically, we will introduce time-variation in the
intercept of the measurement equation, $\xi$, by assuming it is driven
by a hidden Markov process, $\{s_{t}\}$. One may think of the states
as representing different market conditions. We refer to this model
as the Markov-switching Realized GARCH model. We assume that the state
variable, $s_{t}\in\{e_{1},\ldots,e_{N}\}$, follows a stationary
discrete-time hidden Markov chain process on $(\Omega,\mathcal{G},\mathbb{P})$,
with transition probabilities, $\pi_{ij}\equiv\Pr(s_{t+1}=e_{j}|s_{t}=e_{i})$,
$i,j=1,\ldots,N$. So, the Markov-switching Realized GARCH model is
given by (\ref{eq:return}), (\ref{eq:garch}), and a state-dependent
measurement equation:
\begin{equation}
\log x_{t}=\xi_{s_{t}}+\phi\log h_{t}+\delta_{1}z_{t}+\delta_{2}(z_{t}^{2}-1)+\sigma u_{t}.\tag{{\ref{eq:measurement}'}}\label{eq:MSRG-measurement}
\end{equation}
The conventional Realized GARCH model corresponds to the case with
a single state ($N=1$). We can, without loss of generality, represent
the states by the $N$ unit vectors, i.e. let $e_{j}$ be the $j$-th
column of the $N\times N$ identity matrix $I_{N}$. With this convention
we have $\xi_{s_{t}}=\xi^{\prime}s_{t}$, where $\xi\in\mathbb{R}^{N\times1}$
is a vector of parameters with $\xi_{j}\equiv\xi_{(s_{t}=e_{j})}$,
$j=1,\ldots,N$. Note that the parameter $\xi_{j}$ controls the long-run
volatility level within each state because the dynamic properties
of $\log h_{t}$ are given by:
\[
\log h_{t+1}=\omega+\gamma\xi^{\prime}s_{t}+(\beta+\gamma\phi)\log h_{t}+(\tau_{1}+\gamma\delta_{1})z_{t}+(\tau_{2}+\gamma\delta_{2})(z_{t}^{2}-1)+\gamma\sigma u_{t}.
\]
Thus, if $|\beta+\gamma\phi|<1$, then $\log h_{t}$ is mean-reverting
towards $\left(\omega+\gamma\xi_{j}\right)/(1-\beta-\gamma\phi)$
in the $j$-th state, $j=1,\ldots,N$. Thus, the Markov switching
model for $\xi$ is effectively a Markov switching model for the long-run
level of volatility. This highlights the motivations for introducing
time-variation in this parameter. Another advantage of mapping states to the measurement equation is that inference
about states can be simply inferred from the realized measures and returns. Introducing Markov switching in other parameters would greatly complicate the analysis and, in
some cases, make analytical option pricing formulae unobtainable.

\subsection{Risk Neutralization ($\mathbb{Q}$)}

Next, we turn to the risk neutralization as characterized by the pricing
kernel. For this purpose, we generalized the exponentially affine
pricing kernel 
\[
M_{t+1,t}=\frac{\exp[\psi z_{t+1}+\chi u_{t+1}]}{\mathbb{E}_{t}^{\mathbb{P}}[\exp(\psi z_{t+1}+\chi u_{t+1})]},
\]
to be state-dependent. Here $\mathbb{E}_{t}^{\mathbb{P}}(\cdot)\equiv\mathbb{E}^{\mathbb{P}}(\cdot|\mathcal{F}_{t})$.
This kernel was introduced within the Realized GARCH framework by
\citet{HuangWangHansen2017}, where the parameters, $\psi$ and $\chi$,
govern the equity risk premium and the variance risk premium, respectively.
We generalize this pricing kernel by substituting $\psi_{s_{t}}$
for $\psi$ and $\chi_{s_{t}}$ for $\chi$. However, it immediately
follows that the former must be constant, because a no-arbitrage condition
has the implication that $\psi_{s_{t}}=-\lambda$, which rules out
time-variation in $\psi$, see Appendix \ref{eq:psi=00003Dminuslambda}.
Thus, the time-variation across states is confined to that in $\chi_{s_{t}}$,
and our state-dependent pricing kernel is given by:
\begin{equation}
M_{t+1,t}(s_{t+1})=\frac{\exp(\psi z_{t+1}+\chi_{s_{t+1}}u_{t+1})}{\mathbb{E}_{t}^{\mathbb{P}}[\exp(\psi z_{t+1}+\chi_{s_{t+1}}u_{t+1})|s_{t+1}]}.\label{eq:PricingKernelNew}
\end{equation}
Note that the additional conditioning on $s_{t+1}$ is required
due to the additional source of uncertainty induced by the
hidden regime-switching part of the model. The appropriate martingale
condition is, therefore, given by the enlarged filtration: $\mathcal{G}_{t+1}\vee\mathcal{F}_{t}$,
where $\mathcal{G}_{t}=\ensuremath{\sigma}(\{s_{\tau}\},\tau\leq t).$
Naturally, by the law of iterated expectations it follows that the
martingale condition holds with $\mathcal{F}_{t}$ alone, if it holds
for the enlarged filtration. And $\mathcal{F}_{t}$ conveniently does
not require knowledge about the latent state. The dynamic properties
under the risk-neutral measure, $\mathbb{Q}$, are as
stated in the following theorem.
\begin{thm}
\label{thm:Qdynamics}Suppose that returns and realized volatilities
under the $\mathbb{P}$-measure follow the Markov-switching Realized
GARCH model in (\ref{eq:return}), (\ref{eq:garch}), (\ref{eq:MSRG-measurement}),
and the pricing kernel is given by (\ref{eq:PricingKernelNew}). Then, under $\mathbb{Q}$, we have that $z_{t+1}^{*}\equiv z_{t+1}+\lambda$ and
$u_{t+1}^{*}\equiv u_{t+1}-\chi_{s_{t+1}}$ are independent and identically
distributed, $N(0,1)$, and
\begin{eqnarray*}
R_{t+1} & = & r-\tfrac{1}{2}h_{t+1}+\sqrt{h_{t+1}}z_{t+1}^{*},\\
\log h_{t+1} & = & \omega^{*}+\beta\log h_{t}+\gamma\log x_{t}+\tau_{1}^{*}z_{t}^{*}+\tau_{2}(z_{t}^{*2}-1),\\
\log x_{t} & = & \xi_{s_{t}}^{*}+\phi\log h_{t}+\delta_{1}^{*}z_{t}^{*}+\delta_{2}(z_{t}^{*}{}^{2}-1)+\sigma u_{t}^{*},
\end{eqnarray*}
where $\omega^{*}=\omega-\tau_{1}\lambda+\tau_{2}\lambda^{2}$, $\tau_{1}^{*}=\tau_{1}-2\tau_{2}\lambda$,
$\delta_{1}^{*}=\delta_{1}-2\delta_{2}\lambda$ and 
\[
\xi_{s_t}^{*}=\xi_{s_t}-\delta_{1}\lambda+\delta_{2}\lambda^{2}+\sigma \chi_{s_t}.
\]
\end{thm}
We seek to characterize the implications of a state-dependent variance
risk premium parameter, $\chi_{s_{t}}$. To this end, we consider the difference between conditional
expectations of log-variance under $\mathbb{P}$ and under $\mathbb{Q}$,
\begin{equation}
\mathbb{E}_{t}^{\mathbb{Q}}(\log h_{t+2})-\mathbb{E}_{t}^{\mathbb{\mathbb{P}}}(\log h_{t+2})=\ensuremath{-\left(\tau_{1}+\gamma\delta_{1}\right)\lambda+\left(\tau_{2}+\gamma\delta_{2}\right)\lambda^{2}+\gamma\sigma\mathbb{E}^{\mathbb{P}}_t(\chi_{s_{t+1}}),}\label{eq: logVRP}
\end{equation}
which is a logarithmic variant of the variance risk premium. 
This quantity can be decomposed into compensation
for equity risk and additional compensation for volatility risk. The former is given by the first two terms in (\ref{eq: logVRP}), which involve $\lambda$, and the latter
is the last term in (\ref{eq: logVRP}), which is state-dependent. Thus, the parameter $\chi_{s_t}$
governs the variance risk aversion of investors in state $s_{t}$,
where a large value of $\chi$ is associated with a high variance
risk premium. 

Note that the expectation $\mathbb{E}^{\mathbb{P}}_t(\chi_{s_{t+1}})$ can be also expressed as $\sum_{s_{t+1}}P_{t}(s_{t+1})\chi_{s_{t+1}}$,
where $P_{t}(s_{t+1})$ is the notation for the conditional distribution
$\Pr(s_{t+1}|\mathcal{F}_{t})$, over the possible states, $e_{1},\ldots,e_{N}$.
Similarly, we use $P_{t}(s_{t})$ as compact notation for $\Pr(s_{t}|\mathcal{F}_{t})$. Next, we will derive model-based expressions for multi-step ahead forecasts
of $h_{t}$, which defines the, so-called, \emph{variance term structure}. 
\begin{cor}
\label{cor:E(h_=00007Bt+n=00007D|s_t)}The model-implied n-periods
ahead forecast of $h_{t}$ within the Markov-switching Realized GARCH
model under $\mathbb{Q}$ is:
\begin{align}
\mathbb{E}_{t}^{\mathbb{Q}}(h_{t+n}|s_{t}) & =\exp\left(\kappa_n + \rho^{n-1}\log h_{t+1}+\theta_{n}^{\prime}s_{t}\right),\label{eq:E(h_=00007Bt+n=00007D|s_t)}
\end{align}
with $\rho=\beta+\gamma\phi$. The vector $\theta_{n}\in\mathbb{R}^{N\times1}$
is given recursively from,
\[
\theta_{n+1}=\Delta \left( \rho^{n-1}\zeta+\theta_{n}\right),\quad\Delta_{i}(\varphi)\equiv\log\left[\sum_{j=1}^{N}\pi_{ij}\exp(\varphi_{j})\right],
\]
with initial condition $\theta_{1}=\bm{0}$, and $\zeta_j = \omega^{*}+\gamma\xi^{*}_j,\ j=1,\ldots,N $. The intercept term is given by $\kappa_{n}=\sum_{i=0}^{n-2}G(\rho^{i})$, where the function  $G(\cdot)$ is the logarithm of the moment generating function (MGF)
for the random variable $\left(\tau_{1}^{*}+\gamma\delta_{1}^{*}\right)z_{t}^{*}+(\tau_{2}+\gamma\delta_{2})(z_{t}^{*}{}^{2}-1)+\gamma\sigma u_{t}^{*}$. 
\end{cor}
Corollary \ref{cor:E(h_=00007Bt+n=00007D|s_t)} shows that the variance
term  structure, $\mathbb{E}_{t}^{\mathbb{Q}}(h_{t+n}|s_{t})$, for $n=1,2,\ldots$,
depends on the current state $s_{t}$ and the current level of volatility,
$h_{t+1}$, where the latter is $\mathcal{F}_{t}$-measurable. A major
benefit of having a closed-from expression for $\mathbb{E}_{t}^{\mathbb{Q}}(h_{t+n}|s_{t})$
is that it leads to an analytical expression for the model-implied
VIX price. In the present context, the VIX pricing formula is given
by 
\[
{\rm{VIX}}_t = A\times\sqrt{\sum_{n=1}^{22}\mathbb{E}_{t}^{\mathbb{Q}}(h_{t+n})}=A\times\sqrt{\sum_{n=1}^{22}\sum_{j=1}^{N}\mathbb{E}_{t}^{\mathbb{Q}}(h_{t+n}|s_{t})P_{t}(s_{t}=e_{j})},
\]
where $A=100\sqrt{252/22}$ is the annualized factor.

\subsection{Model Estimation by Maximum Likelihood ($\mathbb{P}$)\label{subsec:Model-EstimationP}}

In this section, we discuss the estimation of the Markov-switching Realized
GARCH model based on returns and realized measures alone. In  Section \ref{subsec:JointEstimation}, we will discuss how option prices can be incorporated in the estimation of this model in conjunction with the parameters in the pricing kernel.

The model is relatively simple to estimate by maximum likelihood for
$\{(R_{t},x_{t})\}_{t=1}^{T}$. The likelihood for $(R_{t+1},x_{t+1})$
is given by their density, conditional on $\mathcal{F}_{t}$, $L_{t}(R_{t+1},x_{t+1})=f(R_{t+1},x_{t+1}|\mathcal{F}_{t})$,
which  we can rewrite as 
\begin{eqnarray}
L_{t}(R_{t+1},x_{t+1}) & = & \sum_{s_{t+1}}P_{t}(s_{t+1})L_{t}(R_{t+1},x_{t+1}|s_{t+1})\nonumber \\
 & = & \sum_{s_{t+1}}P_{t}(s_{t+1})L_{t}(R_{t+1}|s_{t+1})L_{t}(x_{t+1}|R_{t+1},s_{t+1})\nonumber \\
 & = & \sum_{s_{t+1}}P_{t}(s_{t+1})L_{t}(R_{t+1})L_{t}(x_{t+1}|R_{t+1},s_{t+1}).\label{eq:LL}
\end{eqnarray}
The last equality uses that the distribution of $R_{t+1}=r+\lambda\sqrt{h_{t+1}}-\tfrac{1}{2}h_{t+1}+\sqrt{h_{t+1}}z_{t+1},$
conditional on $\mathcal{F}_{t}$, does not depend on the state, $s_{t+1}$.
The reason is that $h_{t+1}$ is $\mathcal{F}_{t}$-measurable and
$z_{t+1}$ is independent of $s_{t+1}$, such that $L_{t}(R_{t+1}|s_{t+1})=L_{t}(R_{t+1})$.
This term of the likelihood function is
\[
L_{t}(R_{t+1})=\frac{1}{\sqrt{2\pi h_{t+1}}}\exp\left\{ -\frac{1}{2}\frac{[R_{t+1}-(r+\lambda\sqrt{h_{t+1}}-h_{t+1}/2)]^{2}}{h_{t+1}}\right\} .
\]
Next, the conditional distribution of the realized measure, $L_{t}(x_{t+1}|R_{t+1},s_{t+1})$,
is log-normally distributed. From the measurement equation, (\ref{eq:MSRG-measurement}), and the Gaussian specification, we have
\[
L_{t}(x_{t+1}|R_{t+1},s_{t+1})=\frac{1}{\sqrt{2\pi\sigma^{2}}}\exp\left\{ -\frac{1}{2}\frac{[x_{t+1}-(\xi_{s_{t+1}}+\phi\log h_{t+1}+\delta_{1}z_{t+1}+\delta_{2}(z_{t+1}^{2}-1))]^{2}}{\sigma^{2}}\right\} .
\]
Finally, the remaining term in (\ref{eq:LL}), $P_{t}(s_{t+1})$,
is the conditional state distribution. The relation between $P_{t}(s_{t+1})$
and $P_{t}(s_{t})$ is given by
\begin{align}
P_{t}(s_{t+1}) & =\sum_{s_{t}}P_{t}(s_{t+1}|s_{t})P_{t}(s_{t})=\sum_{s_{t}}P(s_{t+1}|s_{t})P_{t}(s_{t}),\label{eq: pt-1st-1}
\end{align}
where the last equality follows by the transition probabilities being
time-invariant and independent of $\{R_{t},x_{t}\}$. Before we can
evaluate (\ref{eq:LL}), we need to compute $P_{t}(s_{t})$.
From Bayes' Theorem we have,
\[
\Pr(s_{t}|R_{t},x_{t},\mathcal{F}_{t-1})=\frac{\Pr(R_{t},x_{t},s_{t}|\mathcal{F}_{t-1})}{\Pr(R_{t},x_{t}|\mathcal{F}_{t-1})}=\frac{\Pr(R_{t},x_{t}|s_{t},\mathcal{F}_{t-1})\Pr(s_{t}|\mathcal{F}_{t-1})}{\Pr(R_{t},x_{t}|\mathcal{F}_{t-1})},
\]
where the first term is identical to $P_{t}(s_{t})$ and the last
is identical to $\frac{L_{t-1}(R_{t},x_{t}|s_{t})P_{t-1}(s_{t})}{L_{t-1}(R_{t},x_{t})}$.
So, 
\begin{eqnarray}
P_{t}(s_{t}) & = & \frac{L_{t-1}(R_{t},x_{t}|s_{t})P_{t-1}(s_{t})}{L_{t-1}(R_{t},x_{t})}=\frac{L_{t-1}(R_{t},x_{t}|s_{t})P_{t-1}(s_{t})}{\sum_{s_{t}^{\prime}}L_{t-1}(R_{t},x_{t}|s_{t}^{\prime})P_{t-1}(s_{t}^{\prime})}\nonumber \\
 & = & \frac{L_{t-1}(R_{t})L_{t-1}(x_{t}|R_{t},s_{t})P_{t-1}(s_{t})}{\sum_{s_{t}^{\prime}}L_{t-1}(R_{t})L_{t-1}(x_{t}|R_{t},s_{t}^{\prime})P_{t-1}(s_{t}^{\prime})}=\frac{L_{t-1}(x_{t}|R_{t},s_{t})P_{t-1}(s_{t})}{\sum_{s_{t}^{\prime}}L_{t-1}(x_{t}|R_{t},s_{t}^{\prime})P_{t-1}(s_{t}^{\prime})}.\label{eq:P_st-1}
\end{eqnarray}
In the third equality, we used that $L_{t-1}(R_{t}|s_{t})=L_{t-1}(R_{t})$.
By combining (\ref{eq: pt-1st-1}) and (\ref{eq:P_st-1}), we arrive
at the following recursive formula for $P_{t}(s_{t+1})$:
\begin{equation}
P_{t}(s_{t+1})=\frac{\sum_{s_{t}}\pi_{s_{t},s_{t+1}}L_{t-1}(x_{t}|R_{t},s_{t})P_{t-1}(s_{t})}{\sum_{s_{t}^{\prime}}L_{t-1}(x_{t}|R_{t},s_{t}^{\prime})P_{t-1}(s_{t}^{\prime})},\label{eq:PPequ}
\end{equation}
that defines the recursion from $P_{t-1}(s_{t})$ to $P_{t}(s_{t+1})$.
Therefore, given an initial value for $P_{0}\left(s_{1}\right)$,
we can evaluate the likelihood function for $\{(R_{t},x_{t})\}_{t=1}^{T}$.
There are two obvious ways to obtain an initial value for $P_{0}\left(s_{1}\right)$.
One can either model it as an unknown vector (of conditional state
probabilities) to be estimated, or one can assign $P_{0}\left(s_{1}\right)$
to have its stationary distribution, which is given by the eigenvector,
$\pi^{\prime}\Pi=\pi^{\prime}$. In our empirical implementation,
we use the latter.

The recursion for state probabilities, (\ref{eq:PPequ}), depends
on the conditional distribution of realized measures, but does not
depend on the conditional distribution for returns. The reason is
that the state, $s_{t}$, only impacts the intercept in the measurement
equation. The recursive expression for state probabilities, (\ref{eq:PPequ}),
also holds when option prices are included in the analysis and parameters
are estimated by maximization of the total likelihood. However, the
joint estimation will influence all parameter estimates, which will
influence the likelihood for the realized measure, $L_{t-1}(x_{t}|R_{t},s_{t})$.
Thus, the estimated state probabilities based on $\{(R_{t},x_{t})\}_{t=1}^{T}$,
need not be identical to those obtained from joint estimation that
also includes option prices. 

\section{Analytical Approximation for European Call Options}

The properties of the model under the $\mathbb{Q}$-measure were presented
in Theorem \ref{thm:Qdynamics}, and it is evident that there is no
analytical formula for the moment generating function (MGF) of cumulative
return. This complication is not a consequence of the Markov switching
structure because the complication is also present without Markov switching ($N=1$).
Without an analytical formula for the MGF, the conventional path to
closed-form pricing expressions with Fourier inversion is not applicable.
Instead, we follow the method developed in \citet{HuangWangHansen2017}
and derive an analytical approximation based on an Edgeworth expansion
of the density for the cumulative return. 

The European call option price is given by
\[
C_{t}=e^{-r(T-t)}\mathbb{E}_{t}^{\text{\ensuremath{\mathbb{Q}}}}\left(\max\left(S_{T}-K,0\right)\right)=\sum_{s_{t}}P_{t}(s_{t})C_{t}(s_{t}),
\]
where $C_{t}(s_{t})\equiv e^{-r(T-t)}\mathbb{E}_{t}^{\text{\ensuremath{\mathbb{Q}}}}\left(\max\left(S_{T}-K,0\right)|s_{t}\right)$,
$T$ is the date of maturity, $S_{T}$ is the terminal stock price,
and $K$ is the strike price. Let $R_{T}=\log(S_{T}/S_{t})$ be the
future cumulated return, $\mu=\mathbb{E}_{t}^{\text{\ensuremath{\mathbb{Q}}}}(R_{T}|s_{t})$,
and $\sigma^{2}={\rm Var}_{t}^{\mathbb{Q}}\left(R_{T}|s_{t}\right)$, then the expectation can be written as an integral of the standardized
cumulated return $z_{T}=\left(R_{T}-\mu\right)/\sigma$:
\[
C_{t}(s_{t})=e^{-r(T-t)}\int_{-\infty}^{k}\left[S_{t}\exp(\mu-\sigma z)-K\right]\tilde{g}(z)\mathrm{d}z,
\]
where $k=\left(\log\left(S_{t}/K\right)+\mu\right)/\sigma$ and $\tilde{g}(z)$
is the true conditional density function of $-z_{T}$ given $s_{t}$.
Following \citet{JarrowRudd1982}, we apply a second-order Edgeworth
expansion by expanding the density of the standardized cumulated return,
$z_{T}$, using the expression,
\begin{equation}
\text{\ensuremath{\tilde{g}}}(z)\approx\left[1-\frac{\kappa_{3}}{6}H_{3}(z)+\frac{\left(\kappa_{4}-3\right)}{24}H_{4}(z)+\frac{\kappa_{3}^{2}}{72}H_{6}(z)\right]\phi(z).\label{eq:ApproxPDF_zT}
\end{equation}
Here $\phi(z)=(2\pi)^{-\tfrac{1}{2}}\exp(-z^{2}/2$) is the density
of the standard normal distribution, $\kappa_{3}=\mathbb{E}_{t}^{\mathbb{Q}}(z_{T}^{3}|s_{t})$,
$\kappa_{4}=\mathbb{E}_{t}^{\mathbb{Q}}(z_{T}^{4}|s_{t})$, and $H_{n}(z)$
is the $n$-th order Hermite polynomial given by
\[
H_{3}(z)=z^{3}-3z,\qquad H_{4}(z)=z^{4}-6z^{2}+3,\quad\text{and}\quad H_{6}(z)=z^{6}-15z^{4}+45z^{2}-15.
\]
From \citet[proposition 1]{HuangWangHansen2017},
the  price of a European call option is approximated by
\begin{equation}
\ensuremath{C_{t}(s_{t}) \ \approx \ A_{0}+\frac{\kappa_{3}}{6}A_{3}+\frac{\left(\kappa_{4}-3\right)}{24}A_{4}+\frac{\kappa_{3}^{2}}{72}A_{6},}\label{eq:Cappro}
\end{equation}
where
\begin{eqnarray*}
A_{0} & = & S_{t}e^{\delta\sigma}\Phi(d)-Ke^{-r(T-t)}\Phi(d-\sigma),\\
A_{3} & = & S_{t}e^{\delta\sigma}\sigma\left[(2\sigma-d)\phi(d)+\sigma^{2}\Phi(d)\right],\\
A_{4} & = & S_{t}e^{\delta\sigma}\sigma\left[\left(d^{2}-1-3\sigma(d-\sigma)\right)\phi(d)+\sigma^{3}\Phi(d)\right],\\
A_{6} & = & S_{t}e^{\delta\sigma}\sigma\left[\sigma^{5}\Phi(d)+\left(3-6d^{2}+d^{4}+5\sigma\left(d-(d-\sigma)(\sigma d-2)-(d-\sigma)^{3}\right)\right)\phi(d)\right],\\
d & = & \frac{\log\left(S_{t}/K\right)+\mu}{\sigma}+\sigma.
\end{eqnarray*}
Before we can obtain options prices with (\ref{eq:Cappro}), we need
to derive the third and fourth moments of the standardized cumulative
return, $z_{T}$,
\[
\kappa_{3}=\frac{1}{\sigma^{3}}\left[\mathbb{E}_{t}^{\mathbb{Q}}\left(R_{T}^{3}|s_{t}\right)-\mu^{3}\right]-3\frac{\mu}{\sigma},\quad\text{and}\quad\kappa_{4}=\frac{1}{\sigma^{4}}\left[\mathbb{E}_{t}^{\mathbb{Q}}\left(R_{T}^{4}|s_{t}\right)-\mu^{4}\right]-2\frac{\mu}{\sigma}\left(2\kappa_{3}+3\frac{\mu}{\sigma}\right).
\]
The remaining problem is to obtain expressions for $\mathbb{E}_{t}^{\mathbb{Q}}(R_{T}^{3}|s_{t})$
and $\mathbb{E}_{t}^{\mathbb{Q}}(R_{T}^{4}|s_{t})$. These expressions
involve a large number of terms, which are derived 
in Appendix \ref{sec:Terms-for-Analytical}.

In Figure \ref{fig: DensityREG}, we present the approximated density
for $z_{T}$ (under $\mathbb{Q}$) and the true density. The latter
is simulated from the model we estimated in our empirical application,
see Table 2. The model has two states, and the densities are for the
standardized cumulative return, $z_{T}$, over six months in the high-volatility
state. The red solid line is the approximated risk-neutral distribution
and the blue line is the ``true'' risk-neutral distribution for
$z_{T}$. The latter was obtained with 100,000 simulations. The approximation
method largely agrees with the true density.
\begin{figure}[H]
\centering{}\includegraphics[scale=0.9]{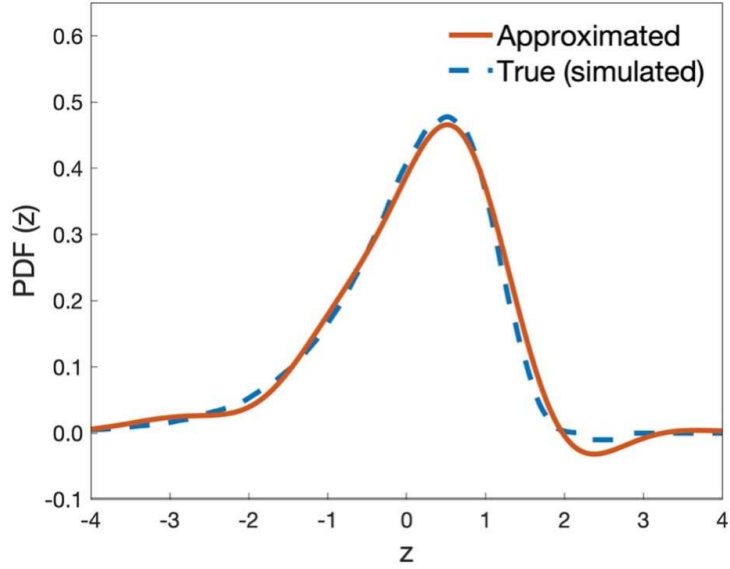}\caption{{\small{}The approximate risk-neutral distribution of standardized
cumulated return, $z_{T}$, over six months for the Markov-switching
Realized GARCH model based on the Edgeworth expansion (red solid line) and
the true density (blue dashed line). The true density is obtained
by 100,000 simulations, using a design based on the estimated parameters
in Table \ref{tab:EstimationTab}.\label{fig: DensityREG}}}
\end{figure}

\section{Model Comparison}

We compare the newly proposed Markov-switching Realized GARCH model
(denote MS-RG) to several existing models, including a range of models
that have documented the benefit of incorporating realized measures
into discrete-time option pricing models. We also include the variant
of the Realized GARCH model by \citet{HuangWangHansen2017} (denote
RG), which is nested in our framework as the case $N=1$. 

\subsection{GARV}

The Generalized Affine Realized Volatility (GARV) model was proposed
by \citet{ChristoffersenFeunouJacobsMeddahi2014}. This model assumes
that the conditional variance for returns, $\bar{h}_{t}$, has two
components. The first, $h_{t}^{R}$, is driven by returns, and the
second, $h_{t}^{{\rm RV}}$, is driven by the realized measure (${\rm RV}$).
This model takes the form:
\begin{eqnarray*}
\ensuremath{R_{t+1}} & = & r+(\lambda-\tfrac{1}{2})\bar{h}_{t+1}+\sqrt{\bar{h}_{t+1}}z_{t+1},\\
\ensuremath{\bar{h}_{t+1}} & = & \xi h_{t+1}^{R}+(1-\xi)h_{t+1}^{{\rm RV}},\\
h_{t+1}^{R} & = & \omega+\beta h_{t}^{R}+\tau_{2}(z_{t}-\tau_{1}\sqrt{\bar{h}_{t}})^{2},\\
h_{t+1}^{{\rm RV}} & = & \kappa+\phi h_{t}^{{\rm RV}}+\delta_{2}(\epsilon_{t}-\delta_{1}\sqrt{\bar{h}_{t}})^{2},\\
{\rm RV}_{t} & = & h_{t}^{{\rm RV}}+\vartheta(\epsilon_{t}^{2}-1-2\delta_{1}\epsilon_{t}\sqrt{\bar{h}_{t}}),
\end{eqnarray*}
where $(z_{t},\epsilon_{t})$ follows a standard bivariate normal
distribution with a correlation of $\rho$. The dynamic properties
under $\mathbb{Q}$ can be obtained through $z_{t}^{*}=z_{t}+\lambda\sqrt{\bar{h}_{t}}$
and $\epsilon_{t}^{*}=\epsilon_{t}+\text{\ensuremath{\chi}}\sqrt{\bar{h}_{t}}$,
where $\left(z_{t}^{*},\epsilon_{t}^{*}\right)$ also follows a standard
bivariate normal distribution with correlation $\rho$. As in the
Realized GARCH model, the parameter $\text{\ensuremath{\chi}}$ is
associated with the variance risk premium, and a positive $\text{\ensuremath{\chi}}$
implies a negative variance risk premium.\footnote{We report $\gamma=\delta_{2}/\vartheta$ instead of $\vartheta$ because
the former can be used to measure the contribution of the realized
information to the volatility process.} Due to the affine structure of the GARV model, a closed-form option
price is available.

\subsection{LHARG}

In addition to GARCH-type models, the availability of high-frequency
data and realized measures has boosted the development of reduced-form
models such as the HAR model. In particular, the HAR model was adapted
by using the leverage function of the Heston-Nandi GARCH as well as
a noncentral gamma distribution (LHARG) to price European
call options. \citet{MajewskiBormettiCorsi2015} provided a general
framework for option pricing with an LHARG model.
\[
R_{t+1}=r+(\lambda-\tfrac{1}{2}){\rm RV}_{t+1}+\sqrt{{\rm RV}_{t+1}}z_{t+1},\quad{\rm RV}_{t+1}|\mathcal{F}_{t}\sim\Gamma\left(\delta,\varTheta_{t},\theta\right),
\]
\[
\ensuremath{\varTheta_{t}\theta=d+\beta_{d}{\rm RV}_{t}^{(d)}+\beta_{w}{\rm RV}_{t}^{(w)}+\beta_{m}{\rm RV}_{t}^{(m)}}\ensuremath{+\alpha_{d}\bar{\ell}_{t}}^{(d)}\ensuremath{+\alpha_{w}\bar{\ell}_{t}}^{(w)}\ensuremath{+\alpha_{m}\bar{\ell}_{t}}^{(m)},
\]
\[
\ensuremath{\begin{aligned}\mathrm{RV}_{t}^{(d)} & =\mathrm{RV}_{t}, & \ell_{t}^{(d)} & =\left(z_{t}^{2}-1-2\gamma z_{t}\sqrt{\mathrm{RV}_{t}}\right),\\
\mathrm{RV}_{t}^{(w)} & =\frac{1}{4}\sum_{i=1}^{4}\mathrm{RV}_{t-i}, & \ell_{t}^{(w)} & =\frac{1}{4}\sum_{i=1}^{4}\left(z_{t-i}^{2}-1-2\gamma z_{t-i}\sqrt{\mathrm{RV}_{t-i}}\right),\\
\mathrm{RV}_{t}^{(m)} & =\frac{1}{17}\sum_{i=5}^{21}\mathrm{RV}_{t-i}, & \ell_{t}^{(m)} & =\frac{1}{17}\sum_{i=5}^{21}\left(z_{t-i}^{2}-1-2\gamma z_{t-i}\sqrt{\mathrm{RV}_{t-i}}\right).
\end{aligned}
}
\]
\citet{MajewskiBormettiCorsi2015} denoted this model by ZM-LHARG
due to the zero-mean leverage function. They found it has the best
option pricing performance because its less-constrained leverage allows
the process to explain a larger fraction of the skewness and kurtosis
observed in real data. The risk-neutral dynamics are given by:
\[
R_{t+1}=r-\tfrac{1}{2}{\rm RV}_{t+1}+\sqrt{{\rm RV}_{t+1}}z_{t+1}^{*},\quad{\rm RV}_{t+1}|\mathcal{F}_{t}\sim\Gamma\left(\delta,\varTheta_{t}^{*},\theta^{*}\right),
\]
where $z_{t}^{*}=z_{t}+\lambda\sqrt{\mathrm{RV}_{t}}$ follows a standard
normal distribution. The risk-neutral parameters are linked to the
physical parameters as follows:
\[
\varTheta_{t}^{*}=\text{\ensuremath{\text{\ensuremath{\chi}}}}\varTheta_{t},\ \theta^{*}=\text{\ensuremath{\chi}}\theta,\ d^{*}=\text{\ensuremath{\chi}}^{2}d,\ \beta_{j}^{*}=\text{\ensuremath{\chi}}^{2}(\beta_{j}+\alpha_{j}\left((\gamma+\lambda)^{2}-\gamma^{2}\right)),\ \alpha_{j}^{*}=\text{\ensuremath{\chi}}^{2}\alpha_{j},\ {\rm for}\ensuremath{\ j\in\{d,w,m\}},
\]
where $\ensuremath{\chi=(1+\theta(\tfrac{1}{2}(\lambda-\tfrac{1}{2})^{2}-v_{0}-\tfrac{1}{8}))^{-\tfrac{1}{2}}}$.
Here, a negative $v_{0}$ is associated with a negative variance risk
premium. \citet{MajewskiBormettiCorsi2015} derive the option-pricing
formula for this model.

\subsection{Heston--Nandi GARCH model}

The last competing model is the Heston--Nandi GARCH model (hereafter
HNG) presented by \citet{HestonNandi2000}. HNG is one of very few
discrete-time volatility models that yield an analytical option pricing
formula. We adopt the variance dependent pricing kernel introduced
in \citet{ChristoffersenHestonJacobs2013} to risk neutralize the
HNG model, which takes the variance premium into account. The dynamic
properties under the physical measures are given by
\begin{eqnarray*}
\ensuremath{R_{t+1}} & = & r+(\lambda-\tfrac{1}{2})h_{t+1}+\sqrt{h_{t+1}}z_{t+1},\\
h_{t+1} & = & \omega+\beta h_{t}+\tau_{2}\left(z_{t}-\tau_{1}\sqrt{h_{t}}\right)^{2}.
\end{eqnarray*}
The corresponding risk-neutral dynamics with the variance-augmented
pricing kernel are given by:
\begin{eqnarray*}
\ensuremath{R_{t+1}} & = & r-\tfrac{1}{2}h_{t+1}^{*}+\sqrt{h_{t+1}^{*}}z_{t+1}^{*},\\
h_{t+1}^{*} & = & \omega^{*}+\beta h_{t}^{*}+\tau_{2}^{*}\left(z_{t}^{*}-\tau_{1}^{*}\sqrt{h_{t}^{*}}\right)^{2},
\end{eqnarray*}
where $z_{t+1}^{*}$ has a standard normal distribution and the risk-neutral
parameters are
\[
h_{t}^{*}=\chi h_{t},\quad\omega^{*}=\chi\omega,\quad\tau_{2}^{*}=\chi^{2}\tau_{1}\quad\tau_{1}^{*}=\left(\lambda+\tau_{1}-\frac{1}{2}\right)\chi^{-1}+\frac{1}{2}.
\]
Once again, the parameter $\chi$ is associated with the variance
risk premium, where $\chi>1$ corresponds to negative variance risk
premium.

\section{Estimation Method \label{subsec:JointEstimation}}

In this section, we turn to model estimation including the pricing
kernel. In Section \ref{subsec:Model-EstimationP} we estimated the
Markov-switching Realized GARCH model from $\{(R_{t},x_{t})\}_{t=1}^{T}$
alone. Now, we will include option prices in estimation,
and simultaneously estimate parameters in the pricing kernel and the
parameters in the Markov-switching Realized GARCH model. We adopt
the quasi log-likelihood for the option prices from \citet{ChristoffersenHestonJacobs2013}
and combine it with the log-likelihood for $\{(R_{t},x_{t})\}_{t=1}^{T}$. The quasi
log-likelihood function for option prices by \citet{ChristoffersenHestonJacobs2013}
assumes that option pricing errors, measured in units of the Black-Scholes
Vega-units, 
\[
e_{i}=\frac{O_{i}^{\mathrm{Model}}-O_{i}^{\mathrm{Market}}}{\nu_{i}^{bs}},\qquad i=1,\ldots,N,
\]
are normally distributed, $e_{i}\sim iidN(0,\sigma_{e}^{2})$. Here,
$O_{i}^{\mathrm{Model}}$ and $O_{i}^{\mathrm{Market}}$ represent
the model-implied option price and the observed market-based option
price, respectively, $\nu_{i}^{bs}$ is the corresponding Black-Scholes
Vega, and $N$ is the total number of option prices in the sample
period. The Vega-weighted pricing error mimics the difference in the
implied volatilities, and $\sigma_{e}^{2}$ denotes the variance of
the Vega-weighted pricing errors. 

Parameters are estimated by maximizing the total quasi log-likelihood
function, $\ell_{\mathrm{Total}}=\ell_{R,x}+\ell_{o}$, where the expression of $\ell_{R,x}$ is given in Section \ref{subsec:Model-EstimationP}, and $\ell_{o}$ is 
\begin{eqnarray*}
\ell_{o} & = & \left\{ -\frac{N}{2}\log(2\pi)-\frac{1}{2}\sum_{i=1}^{N}\log(\sigma_{e}^{2})-\sum_{i=1}^{N}\frac{1}{2\sigma_{e}^{2}}\left(O_{i}^{\mathrm{Model}}-O_{i}^{\mathrm{Market}})/\nu_{i}^{bs}\right)^{2}\right\} \times\frac{T}{N}.
\end{eqnarray*}
In the inclusion of the quasi log-likelihood for option prices, we
follow \citet{ChristoffersenJacobsOrnthanalai2012}, \citet{Ornthanalai2014},
and \citet{HuangTongWang2019}, and scale $\ell_{o}$ by $\tfrac{T}{N}$.
This amounts to an adjustment for the imbalance between the number
of observed option prices and the number of observation of $(R_{t},x_{t})$,
which serves to prevent that parameter estimation is largely dominated
by the option data.

\section{Empirical Results}

\subsection{Data}

Our empirical analysis is based on close-to-close log-returns for
the S\&P 500 index and the panel of SPX option prices. We use the
realized kernel estimator, by \citet{BNHLS:2008}, as our realized
measure of volatility, where the realized kernel is implemented with
the Parzen kernel.\footnote{The S\&P 500 index was collected from Yahoo Finance. The realized measure was
obtained from the Realized Library of Oxford-Man institute for the
years 2000--2019. Before 2000 we used high-frequency data on S\&P
500 futures prices (front-month continuous) from TickData to construct the realized kernels.}

Our sample spans the period from January 1990 to December 2019, which
has 7,559 trading days. The option prices are assembled from two different
databases. Option price data for the first six years are based on
the Optsum data from the CBOE DataShop, whereas option prices, 1996
or later, are based on the OptionMetrics data. We use out-of-the-money
put and call options with positive trading volume and with maturity
between two weeks and six months and we apply the filters proposed
by \citet{BakshiCaoChen1997}. We only use Wednesday option prices,
as is common in this literature. For each maturity, we retain the
three strike prices with the highest liquidity (as defined by daily
trading volume).\footnote{The same type of inclusion criteria were used in \citet{ChristoffersenFeunouJacobsMeddahi2014}
and \citet{HuangWangHansen2017}, who retained the six most liquid
strike prices with maturities between two weeks and six months.} This results in a total of 32,024 option prices.

Table \ref{tab:Summary} reports descriptive statistics for S\&P 500
returns, realized kernels, and CBOE VIX in Panel A. The S\&P 500 returns
exhibit a small negative skewness and a high level of kurtosis. The
realized kernel and the VIX are both positively skewed and leptokurtic.
The standard deviation of returns is, at 17.437\%, substantially smaller
than the average option-implied volatility, at 19.147\%. Their difference
reflects the (average) negative variance risk premium. Panel B of
Table \ref{tab:Summary} provides an extensive summary of the number
of contracts, average prices, and the average implied volatility within
each subcategory in our option data set. Following \citet{ChristoffersenFeunouJacobsMeddahi2014},
the out-of-the-money put options are converted to in-the-money call
options using put-call parity, and the Black-Scholes delta is used
to measure the moneyness of options. The option prices included in
our analysis are all out-of-the-money options. So, options with deltas
larger than 0.5 are out-of-the-money put options, and options with deltas
less than 0.5 are out-of-the-money call options. In the upper part
of panel B, it can be seen that deep out-of-the-money put options
(deltas above 0.7) are relatively expensive compared with out-of-the-money
calls, which reflects the well-known volatility smirk when implied
volatility is plotted against moneyness. The middle part of the panel
B summarizes features of the option prices when sorted by maturity,
and the implied volatility term structure is roughly flat on average.
The bottom of Panel B sorts the data by the volatility state as measured
by contemporary VIX level and, as expected, the implied volatility
increases in VIX.
\begin{table}
\caption{Summary Statistics}

\begin{centering}
\vspace{0.2cm}
\begin{footnotesize}
\begin{tabularx}{\textwidth}{p{3.5cm}YYYYYYY}
\toprule
\midrule
         \\
    \multicolumn{7}{l}{{\it A: S\&P 500 returns, Realized Kernels and CBOE VIX}} \\[6pt]
       &   & Mean(\%) & Std(\%) & Skewness & Kurtosis & Obs. \\[3pt]
    \multicolumn{2}{l}{Returns (annualized)} & 7.377 & 17.437 & -0.267 & 11.851 & 7,559 \\
    \multicolumn{2}{l}{Realized Kernels (annualized)}  & 12.050 & 8.335 & 3.415 & 23.948 & 7,559 \\
    \multicolumn{2}{l}{CBOE VIX}  & 19.147 & 7.725 & 2.132 & 10.955 & 7,559 \\
          &       &       &       &       &  \\
    \\
    \multicolumn{7}{l}{{\it B: SPX Option Price Data}} \\[6pt]
        &  &    \multicolumn{2}{>{\hsize=\dimexpr2\hsize+2\tabcolsep+\arrayrulewidth\relax}c}
        {Implied Volatility (\%) }
        &            \multicolumn{2}{>{\hsize=\dimexpr2\hsize+2\tabcolsep+\arrayrulewidth\relax}c}
        {Average price (\$) }
      & Obs. \\[3pt]
    All options & & \multicolumn{2}{c}{17.53} & \multicolumn{2}{c}{84.24} & 32,024 \\
         & &       &       &       &       &  \\
    \multicolumn{7}{l}{\it Partitioned by moneyness } \\[2pt]
    Delta$<$0.3      &       & \multicolumn{2}{c}{13.05} & \multicolumn{2}{c}{10.73} & 4,475 \\
    0.3$\leq$Delta$<$0.4      &       & \multicolumn{2}{c}{13.93} & \multicolumn{2}{c}{19.38} & 2,464 \\
    0.4$\leq$Delta$<$0.5     &      & \multicolumn{2}{c}{15.07} & \multicolumn{2}{c}{30.81} & 3,093 \\
    0.5$\leq$Delta$<$0.6      &        & \multicolumn{2}{c}{17.09} & \multicolumn{2}{c}{46.38} & 4,449 \\
    0.6$\leq$Delta$<$0.7      &       & \multicolumn{2}{c}{18.16} & \multicolumn{2}{c}{64.92} & 4,178 \\
    0.7$\leq$Delta    &        & \multicolumn{2}{c}{20.22} & \multicolumn{2}{c}{151.81} & 13,365 \\
&          &       &       &       &       &  \\
    \multicolumn{7}{l}{\it Partitioned by maturity} \\[2pt]
     DTM$<$30 &        & \multicolumn{2}{c}{15.48} & \multicolumn{2}{c}{45.80} & 8,255 \\
     30$\leq$DTM$<$60 &        & \multicolumn{2}{c}{17.15} & \multicolumn{2}{c}{75.67} & 8,506 \\
     60$\leq$DTM$<$90 &        & \multicolumn{2}{c}{18.74} & \multicolumn{2}{c}{90.21} & 5,771 \\
    90$\leq$DTM$<$120 &        & \multicolumn{2}{c}{19.19} & \multicolumn{2}{c}{113.89} & 4,338 \\
    120$\leq$DTM$<$150 &        & \multicolumn{2}{c}{18.74} & \multicolumn{2}{c}{125.86} & 2,642 \\
     150$\leq$DTM &       & \multicolumn{2}{c}{18.65} & \multicolumn{2}{c}{130.84} & 2,512 \\
          &       & &      &       &       &  \\
    \multicolumn{7}{l}{\it Partitioned by the level of VIX} \\[2pt] 
     VIX$<$15     &      & \multicolumn{2}{c}{13.04} & \multicolumn{2}{c}{80.51} & 14,746 \\
     15$\leq$VIX$<$20     &        & \multicolumn{2}{c}{17.59} & \multicolumn{2}{c}{88.61} & 8,792 \\
     20$\leq$VIX$<$25     &        & \multicolumn{2}{c}{21.91} & \multicolumn{2}{c}{86.30} & 4,851 \\
     25$\leq$VIX$<$30     &      & \multicolumn{2}{c}{25.74} & \multicolumn{2}{c}{88.97} & 1,965 \\
     30$\leq$VIX$<$35    &        & \multicolumn{2}{c}{29.85} & \multicolumn{2}{c}{86.80} & 856 \\
     35$\leq$VIX     &       & \multicolumn{2}{c}{39.37} & \multicolumn{2}{c}{78.06} & 814 \\
\\[0.0cm]
\\[-0.5cm]
\midrule
\bottomrule
\end{tabularx}
\end{footnotesize}

\par\end{centering}
{\small{}Note: Summary statistics for close-to-close S\&P 500 index
returns, realized kernels (in square root), CBOE VIX, and SPX option
prices from January 1990 to December 2019. The reported statistics
for S\&P 500, realized kernels, and VIX index include the sample mean
(Mean), standard deviation (Std), skewness (Skew), kurtosis (Kurt), and the
number of observations (Obs). Option prices are based on closing prices
of out-of-the-money call and put options. Out-of-the-money put options
are converted to in-the-money call options using put-call parity.
We report the average Black-Scholes implied volatility (IV), average
price, and the number of option prices for different partitions of
option prices. ``Moneyness'' is defined by the Black-Scholes delta.
DTM denotes the number of calendar days to maturity. Data sources:
S\&P 500 returns from Yahoo Finance; VIX from CBOE's website; Realized kernels
from TickData (1990-1999) and Realized Library of Oxford-Man institute
(2000-2019); Option prices from Optsum data (1990--1995) and OptionMetrics
(1996--2019).\label{tab:Summary}}{\small\par}
\end{table}

\subsection{Parameter Estimation}

We present results for the Markov-switching Realized GARCH model with
two states ($N=2$).\footnote{The number of states could be chosen by a suitable statistical criterion.
Nevertheless, a model with two states will likely be preferred in
many applications because additional states add many parameters to
the model, and a two-state model already adds a high degree of flexibility
to the model structure.} In Table \ref{tab:EstimationTab}, we present the estimation results
for each of the models. Parameters are estimated using the sample
from January 1990 to December 2019. The estimated parameters
are given along with their robust standard errors in brackets. We
also report implied logarithmically transformed expectation of $h_{t}$,
and the implied persistence of volatility dynamics, $\pi^{\ensuremath{\mathbb{P}}}$
and $\pi^{\ensuremath{\mathbb{Q}}}$, under the physical and risk-neutral
measures, respectively. The values of the maximized log-likelihood,
$\ell_{\mathrm{Total}}$, are provided. The maximized log-likelihoods are not comparable
unless the underlying models describe the same data, which is not the case for all models. The first
four models are GARCH-type models, which largely share a common notation for their parameters. To conserve space, we present the estimates for $\sigma$
(for MS-RG and RG) and $\rho$ (for GARV) on the same line. The last
model is the LHARG model, which has a different set of parameters and these are labelled in a separate column.\footnote{The logarithm of unconditional variances are estimated instead of the
intercepts in the variance equations, which are then implied from the unconditional
variance formulas.}
\begin{table}
\caption{Estimation Results}

\begin{centering}
\vspace{0.2cm}
\begin{footnotesize}
\begin{tabularx}{\textwidth}{XYYYYYYY}
\toprule
\midrule
    Model & MS-RG   & RG   & GARV   & HNG  &  & LHARG   \\
    \midrule
\\[-0.275cm]
    $\lambda$  & 0.0354 & 0.0359 & 0.2254 & 2.9196 & $\lambda$ & 2.8969 \\
          & \textit{(0.0103)} & \textit{(0.0092)} & \textit{(0.0247)} & \textit{(0.7198)} &       & \textit{(0.3353)} \\
\\[-0.275cm]    
    $\beta$   & 0.8435 & 0.8741 & 0.9776 & 0.6601 & {$\theta$} & 4.14E-05 \\
          & \textit{(0.0002)} & \textit{(0.0002)} & \textit{(0.0033)} & \textit{(0.0231)} &       & \textit{(3.97E-06)} \\
\\[-0.275cm]
    $\tau_1$   & -0.1520 & -0.1157 & 249.89  & 468.59  & {$\delta$} & 1.2182 \\
          & \textit{(0.0007)} & \textit{(0.0011)} & \textit{(22.88)}  & \textit{(31.17)}  &       & \textit{(0.0846)} \\
\\[-0.275cm]
    $\tau_2$  & -0.0080 & 0.0030 & 3.16E-07 & 1.49E-06 & {$\beta_d$} & 0.2060 \\
          & \textit{(0.0009)} & \textit{(0.0017)} &  \textit{(1.63E-08)} & \textit{(7.55E-08)} &       & \textit{(0.0247)} \\
\\[-0.275cm]
    $\gamma$ & 0.1278 & 0.1083 & 0.3056 &       & {$ \beta_w$} & 0.2456 \\
          & \textit{(0.0016)} & \textit{(0.0035)} & \textit{(0.0156)} &       &       & \textit{(0.0192)} \\
\\[-0.275cm]
    $\phi$   & 0.9930 & 0.9952 & 9.23E-08 &       & {$\beta_m$} & 0.2944 \\
          & \textit{(0.0035)} & \textit{(0.0014)} & \textit{(1.67E-03)} &       &       & \textit{(0.0178)} \\
\\[-0.275cm]
    $\delta_1$     & -0.1733 & -0.1641 & 629.75 &       & {$ \alpha_d$} & 4.70E-06 \\
          & \textit{(0.0091)} & \textit{(0.0091)} & \textit{(0.72)} &       &       & \textit{(2.59E-07)} \\
\\[-0.275cm]
    $\delta_2$    & 0.1548 & 0.1707 & 2.47E-06 &       & {$ \alpha_w$} & 8.22E-06 \\
          & \textit{(0.0094)} & \textit{(0.0129)} & \textit{(9.59E-11)}  &       &       & \textit{(3.55E-07)} \\
\\[-0.275cm]
    $\sigma, \rho$ & 0.6306 & 0.6249 & 0.3961 &       & {$ \alpha_m$} & 1.74E-10 \\
          & \textit{(0.0066)} & \textit{(0.0065)} & \textit{(0.0153)} &       &       & \textit{(2.59E-07)} \\
\\[-0.275cm]
    $\xi$    &       & -0.8212 & 0.0584 &       & {$\gamma$} & 409.56 \\
          &       & \textit{(0.0111)} & \textit{(0.0036)} &       &       & \textit{(21.53)} \\
\\[-0.275cm]
    $\xi_{\rm{Low}}$ & -0.8767 &       &       &       &       &  \\
          & \textit{(0.0454)} &       &       &       &       &  \\
\\[-0.275cm]    
    $\xi_{\rm{High}}$ & -0.7788 &       &       &       &       &  \\
          & \textit{(0.0349)} &       &       &       &       &  \\
\\[-0.275cm]
    $\chi$     &       & 0.0519 & 6.5567 & 1.1512 & $\chi$ & 1.0947 \\
          &       & \textit{(0.0129)} & \textit{(0.6418)} & \textit{(0.0222)} &       & \textit{(0.0101)} \\
\\[-0.275cm]    
    $\chi_{\rm{Low}}$  & -0.0052 &       &       &       &       &  \\
          & \textit{(0.0022)} &       &       &       &       &  \\
\\[-0.275cm]
    $\chi_{\rm{High}}$ & 0.4586 &       &       &       &       &  \\
          & \textit{(0.0109)} &       &       &       &       &  \\
\\[-0.275cm]
    $\pi_{{\rm Low|Low}}$   & 0.9996 &       &       &       &       &  \\
          & \textit{(0.0072)} &       &       &       &       &  \\
\\[-0.275cm]    
    $\pi_{{\rm High|High}}$   & 0.9948 &       &       &       &       &  \\
          & \textit{(0.0025)} &       &       &       &       &  \\
          &       &       &       &       &       &  \\
\\[-0.275cm]
    $\sigma_e\times 100$ & 2.4155 & 2.8190 & 2.9756 & 3.4506 & $\sigma_e\times 100$ & 3.1869 \\
\\[-0.275cm]
\\[-0.275cm]    
    $\log {\mathbb E}(h_t)$ & -9.3672 & -9.0763 & -9.0241 & -9.0665 & $\log {\mathbb E}(h_t)$ & -9.4541 \\   
    $\pi^{\mathbb P}$ & 0.9704 & 0.9818 & 0.9974 & 0.9871 & $\pi^{\mathbb P}$ & 0.7460 \\
    $\pi^{\mathbb Q}$ & 0.9704 & 0.9818 & 0.9974 & 0.9912  & $\pi^{\mathbb Q}$ & 0.8503 \\
          &       &       &       &       &       &  \\
       $\ell^{\mathbb P}$ & 17980 & 17964 & 84469 & 24848 & $\ell^{\mathbb P}$ & 89553 \\
        $\ell^{\mathbb Q}$ & 17418 & 16251 & 15833 & 14723 & $\ell^{\mathbb Q}$ & 15329 \\ 
    $\ell_{\mathrm{Total}}$ & 35398 & 34215 & 100302 & 39571 & $\ell_{\mathrm{Total}}$ & 104882 \\
\\[-0.275cm]
\midrule
\bottomrule
\end{tabularx}
\end{footnotesize}
\par\end{centering}
{\small{}Note: Estimation results for the full sample period (January
1990 to December 2019). The first row refers to model specification.
Parameter estimates are reported with robust standard errors (in parentheses),
$\pi^{\mathbb{P}}$ and $\pi^{\mathbb{Q}}$ refer to the volatility
persistence under $\mathbb{P}$ and $\mathbb{Q}$, respectively. The
value of the log-likelihood function is reported at the bottom of
the table where $\ell^{\mathbb{P}}=\ell_{R,x}$ are the terms of the
log-likelihood related to returns and realized measures and $\ell^{\mathbb{Q}}=\ell_{o}$
is the part of the log-likelihood related to option prices.\label{tab:EstimationTab}}{\small\par}
\end{table}

The estimated models are consistent with several stylized facts in
the related literature. First, the estimated models imply a highly persistent
volatility process under both physical and risk-neutral measures.
The only exception is the LHARG model which estimates the volatility
to be less persistent, in particular under the physical measure. The
likely explanation is that the LHARG model treats the noisy realized
measure as the true underlying volatility. It is well known that the
measurement errors in the realized measure will induce a downwards
bias in autocorrelations, see \citet{Hansen2014a}, and this may explain
the lower estimate of persistence in this model. Second, all of the
models estimate the equity premium parameter to be positive, $\lambda>0$,
and significant. Third, the values of the variance risk parameter
($\chi$) all indicate a significant negative variance risk premium\footnote{Positive $\chi$ for MS-RG, RG, GARV models, and $\chi$ greater than one for HNG, LHARG models corresponds to a negative variance risk
premium.} and higher risk-neutral volatility than their physical counterparts.
Fourth, the models find the leverage effect, as indicated by $\tau_{1}$
(and $\gamma$ for LHARG), to be significant. Fifth, the two Realized
GARCH models (columns 1 and 2), estimate the parameter, $\gamma$,
to be positive and significant. This parameter measures the contribution
of the realized volatility measure in describing the volatility dynamics,
and the realized measures are found to be an important predictor.
The estimate of $\phi$ is close to one, which implies that the realized
measure is proportional to the conditional variance. This reinforces
the label \emph{measurement equation} for (\ref{eq:measurement})
and (\ref{eq:MSRG-measurement}). Compared with other models in Table
\ref{tab:EstimationTab}, the two Realized GARCH models deliver the
smallest Vega-weighted pricing error $\sigma_{e}$. 

The estimation results for the Markov-switching Realized GARCH model
(MS-RG), proposed in this paper, are quite interesting. First, the
state-dependent intercept $\xi$ does improve the empirical fit of
the model, as illustrated by the increased value of the log-likelihood
for the ``physical'' variables $\ell^{\mathbb{P}}$. The parameter
$\xi$ controls the long-run volatility level within each state. For
this reason, we adopt labels ``Low'' and ``High'' to denote the
low- and high-volatility states, respectively. Second, we also find
the estimated variance risk premium parameter, $\chi$, to be state-dependent. In the high-volatility state, $\chi_{{\rm High}}$ is estimated
to be 0.4586, and is highly significant. This estimate translates
to a large negative variance risk premium (in absolute value). In
the low-volatility state, we estimate $\chi_{{\rm Low}}$ to be a small negative value,
which suggests that investors have a slight appetite for variance risk in
this state, \textit{i.e.} a positive variance risk premium. The difference
between these two estimates reinforces the value of introducing a
state-dependent pricing kernel. The estimate of $\chi$ in 
single-state RG model is 0.0519, which falls between $\chi_{{\rm High}}$
and $\chi_{{\rm Low}}$. Third, the estimated transition probabilities
within the same state ($\pi_{{\rm High|High}}$ and $\pi_{{\rm Low|Low}}$)
are both near one, which implies the time-varying process of the hidden states is highly persistent. Fourth,
although we focus on the model with two states, the improvements
relative to the benchmark model are impressive. The MS-RG model lowers
the Vega-weighted pricing error $\sigma_{e}$ by 14.3\% relative to the single-state RG model.

\begin{sidewaysfigure}
\centering{}\includegraphics[scale=0.6]{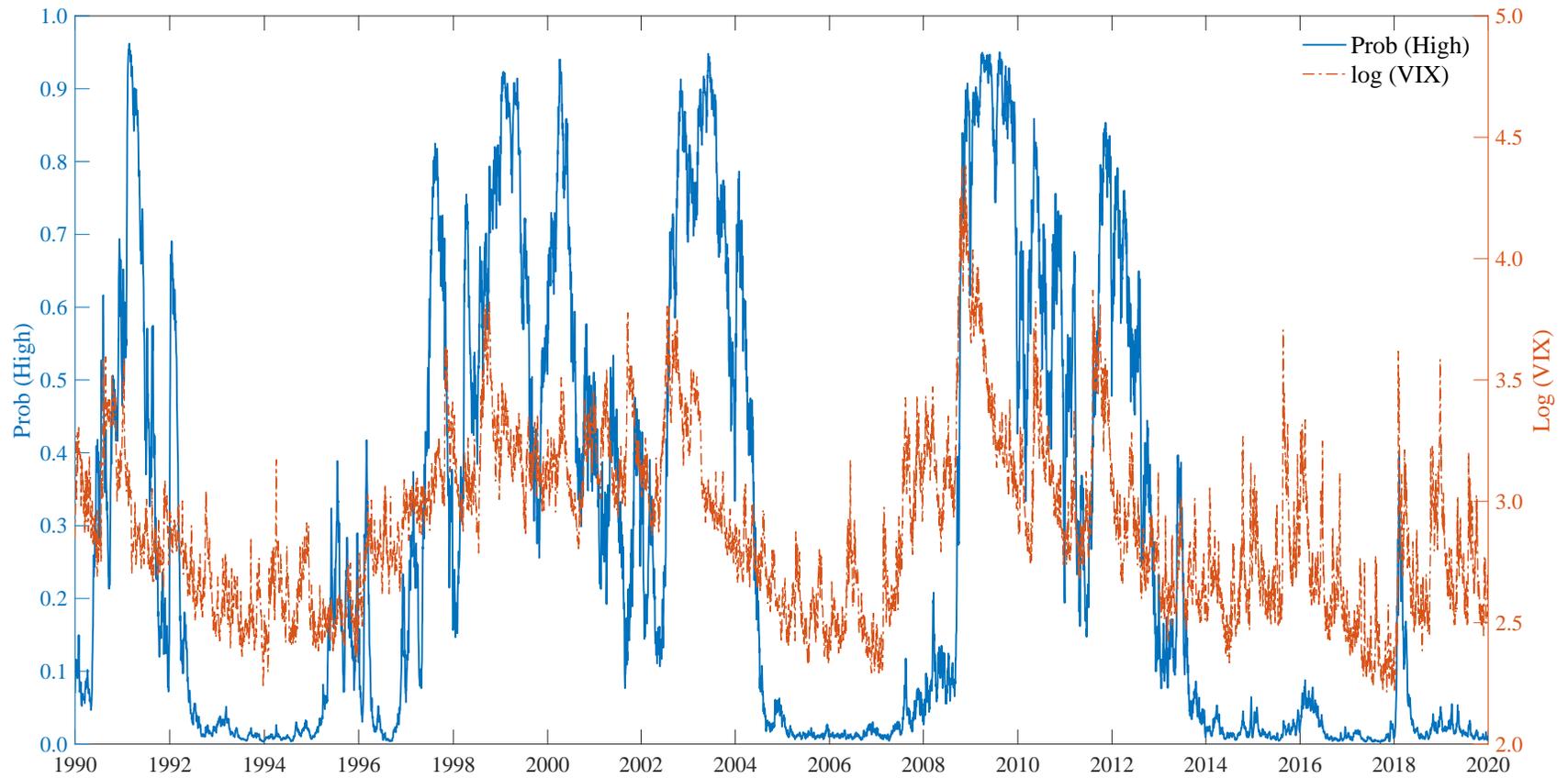}\caption{{\small{}This figure presents the time series of conditional probabilities
in ``High''-volatility state ${\rm Prob}_{t}({\rm High})$ (blue
solid line), and log (VIX) (red dashed line). The correlation between these two
series is 56.89\%.\label{fig:Vertheta}}}
\end{sidewaysfigure}

In Figure \ref{fig:Vertheta}, we present the time series of conditional
state probabilities along with the VIX index. The blue solid line denotes
the estimated value of $P_{t}(s_{t}=\text{\textquotedblleft}\text{High}\text{\textquotedblright})$,
and the red dashed line is the logarithmically transformed VIX index. The
\textquotedblleft High\textquotedblright{} volatility state is also
the state with the highest variance risk premium (in absolute value), so a large value
of $P_{t}(s_{t}=\text{\textquotedblleft}\text{High}\text{\textquotedblright})$
is a period where investors have relatively high variance risk aversion.
There is a great deal of variation in $P_{t}(s_{t}=\text{\textquotedblleft}\text{High}\text{\textquotedblright})$
over time and the process is very persistent because $\pi_{ii}$ is
estimated to be close to one. There are several period, where $P_{t}(s_{t}=\text{\textquotedblleft}\text{Low}\text{\textquotedblright})=1-P_{t}(s_{t}=\text{\textquotedblleft}\text{High}\text{\textquotedblright})\simeq1$,
including most of the time during the year, 1993-1995, 2004-2007,
and 2014-2017. These are times where investors have an appetite for
variance risk (or very low volatility risk aversion). Investors demand
additional compensation for taking on variance risk in the ``High'' volatility 
state. The onset of these periods often coincides with large jumps
in the VIX index, such as those seen around the time of the Asian
Crisis in 1997, the bursting of the dot-com bubble and corporate scandals
in early 2000s, the global financial crises, and the Euro crisis.
A large component of the VIX is, according to \citet{BekaertHoerovaDuca2013},
driven by factors that relate to time-varying risk-aversion. The commonality
between $P_{t}(s_{t}=\text{\textquotedblleft}\text{High}\text{\textquotedblright})$
and the VIX supports this view, and the clear pattern that large upwards
jumps in these two series tend to coincide.

\subsection{Option Pricing Performance}

In this section, we turn to the models' ability to price options,
first in-sample and then out-of-sample. We follow the existing literature
and convert option prices to their corresponding implied volatilities,
as defined by the Black--Scholes formula. We evaluated each of the
models by comparing observed implied volatility, denoted by $\mathrm{IV}^{\mathrm{Market}}$,
to the corresponding model-based implied volatility, denoted $\mathrm{IV}^{\mathrm{Model}}$.
The criterion used in our evaluation is the root mean squared error
(RMSE),
\[
{\rm RMSE_{IV}}=\sqrt{\frac{1}{N}\text{\ensuremath{\sum_{i=1}^{N}\left(\mathrm{IV}_{i}^{\mathrm{Model}}-\mathrm{IV}_{i}^{\mathrm{Market}}\right)^{2}}}}\times100,
\]
where $i=1,\ldots,N$ indexes each of the option prices (converted
to implied volatilities) that were included in the sample period.

\subsubsection{In-sample Option Pricing}

In Table \ref{tab:OptionPricing} we report the in-sample performance
for option pricing for each of the estimated models. The RMSEs are
reported in the first row and we find the MR-RG to have the smallest
RMSE followed by the RG. The HNG has the largest average RMSE. The
percentage reduction in RMSE of the MS-RG model relative
to each of the alternative models ranges from 15.3\% to 30.8\%. 

In order to investigate if the improvements by the MS-RG model are
seen for options with specific characteristics or are found across
the board, we repeat the comparisons after sorting the options by
moneyness, time to maturity, and the contemporaneous level of the
VIX.  

Sorting by moneyness can cast light on the models' ability to generate
sufficient leverage effect. The MS-RG model has the smallest RMSE in all subcategories. Relative to the single-state RG model, the two-state MS-RG model performs  particularly well at pricing deep out-of-the-money call options (Delta$<$0.3), with the reduction in RMSE up to 21\%.

Maturity is related to the models' ability to explain volatility dynamics
over longer time spans. The RMSEs of the MS-RG model are fairly
uniform along this dimension and the MS-RG model also has the smallest
RMSE across all subcategories. The RMSE of the MS-RG is a tad higher
at the longest maturity compared to shorter maturities, but its RMSE
is still substantially smaller than that of all competitors.

Finally, the comparisons across levels of the volatility index will
illustrate the models' ability to generate a proper variance risk
premium at different levels of volatility. Along this dimension, we
also find that the MS-RG model has the smallest RMSE in all subcategories.
The RMSEs are roughly proportional to the level of VIX, which is what
one would expect if the distribution of pricing errors, when measured
in percentage, is relatively homogeneous across subcategories.
\begin{table}
\caption{Option Pricing Performance (In-sample)}

\begin{centering}
\vspace{0.2cm}
\begin{footnotesize}%
\begin{tabularx}{\textwidth}{p{3cm}YYYYYYY}
\toprule
\midrule
    Model & MS-RG   & RG   & GARV   & LHARG  & HNG  \\
    \midrule
          &       &       &       &       &        \\
    Total ${\rm RMSE_{IV}}$ & {2.4190} & {2.8564} & {2.9597} & {3.1633} & {3.4955} \\
          &       &       &       &       &        \\
    \multicolumn{6}{l}{\it Partitioned by moneyness } \\[3pt]
     {Delta$<$0.3     } & {2.4054} & {3.0533} & {2.9437} & {3.0871} & {3.5852} \\
    {0.3$\leq$Delta$<$0.4     } & {2.3072} & {2.7440} & {2.6584} & {3.2202} & {3.4494} \\
    {0.4$\leq$Delta$<$0.5    } & {2.2396} & {2.5964} & {2.4270} & {2.9568} & {3.2505} \\
    {0.5$\leq$Delta$<$0.6     } & {2.5395} & {2.9568} & {2.7650} & {3.3448} & {3.4288} \\
    {0.6$\leq$Delta$<$0.7     } & {2.6245} & {2.9730} & {2.9863} & {3.3649} & {3.5866} \\
    {0.7$\leq$Delta   } & {2.3738} & {2.7942} & {3.1830} & {3.0934} & {3.5230} \\
           &       &       &       &       &       \\
    \multicolumn{6}{l}{\it Partitioned by maturity} \\[3pt]
       { DTM$<$30} & {2.4658} & {2.7672} & {2.8327} & {2.7448} & {3.6105} \\
    { 30$\leq$DTM$<$60} & {2.3509} & {2.8310} & {2.9129} & {3.0095} & {3.4881} \\
    { 60$\leq$DTM$<$90} & {2.3949} & {2.9086} & {3.1281} & {3.4383} & {3.5600} \\
    {90$\leq$DTM$<$120} & {2.3609} & {2.8741} & {2.8234} & {3.3095} & {3.3103} \\
    {120$\leq$DTM$<$150} & {2.3939} & {2.8091} & {3.1161} & {3.4983} & {3.3422} \\
    {150$\leq$DTM} & {2.6617} & {3.1234} & {3.1910} & {3.6585} & {3.4487} \\
          &       &       &       &         &  \\
    \multicolumn{6}{l}{\it Partitioned by the level of VIX} \\[3pt]
    { VIX$<$15    } & {1.5403} & {1.8446} & {1.6880} & {1.8466} & {2.3889} \\
    { 15$\leq$VIX$<$20    } & {2.4148} & {2.5489} & {2.4099} & {2.4991} & {3.2568} \\
    { 20$\leq$VIX$<$25    } & {2.9759} & {3.6254} & {3.9540} & {4.1993} & {4.4586} \\
    { 25$\leq$VIX$<$30    } & {3.0672} & {4.0626} & {4.8481} & {5.0164} & {4.7876} \\
    { 30$\leq$VIX$<$35   } & {3.4921} & {4.8279} & {6.1206} & {5.9721} & {5.4262} \\
    { 35$\leq$VIX    } & {6.0232} & {6.8771} & {6.5122} & {7.9058} & {7.6930} \\
\\[0.0cm]
\\[-0.5cm]
\midrule
\bottomrule
\end{tabularx}
\end{footnotesize}
\par\end{centering}
{\small{}Note: This table reports the in-sample option pricing performance
for each model in Table \ref{tab:EstimationTab} (January 1990 to
December 2019). We evaluate the model's option pricing ability through
the root of mean square errors of implied volatility (${\rm RMSE_{IV}}$).
We summarize the results by option moneyness, maturity and market
VIX level. Moneyness is measured by Delta computed from the Black-Scholes
model. DTM denotes the number of calendar days to maturity.\label{tab:OptionPricing}}{\small\par}
\end{table}
\begin{table}
\caption{Option Pricing Performance (Out-of-sample)}

\begin{centering}
\vspace{0.2cm}
\begin{footnotesize}%
\begin{tabularx}{\textwidth}{p{3.1cm}YYYYYYY}
\toprule
\midrule
    Model & MS-RG   & RG   & GARV   & LHARG  & HNG  \\
    \midrule
          &       &       &       &       &        \\
    Total ${\rm RMSE_{IV}}$  & 2.4835 & 2.9903 & 3.2957 & 3.2960 & 3.7891 \\
          &       &       &       &       &        \\
    \multicolumn{6}{l}{\it Partitioned by moneyness } \\[3pt]
   {Delta$<$0.3     } & 2.5241 & 3.6791 & 4.0796 & 4.1773 & 5.0360 \\
   {0.3$\leq$Delta$<$0.4     } & 2.3688 & 3.1397 & 3.8544 & 3.9975 & 4.4769 \\
   {0.4$\leq$Delta$<$0.5    } & 2.1955 & 2.7655 & 3.4133 & 3.4129 & 3.9895 \\
   {0.5$\leq$Delta$<$0.6     } & 2.5905 & 2.8695 & 3.5277 & 3.6684 & 3.8401 \\
   {0.6$\leq$Delta$<$0.7     } & 2.5965 & 2.7067 & 3.3051 & 3.3755 & 3.6698 \\
   {0.7$\leq$Delta   } & 2.4840 & 2.8429 & 2.7337 & 2.5642 & 3.0452 \\
           &       &       &       &       &       \\
    \multicolumn{6}{l}{\it Partitioned by maturity} \\[3pt]
   { DTM$<$30} & 2.4336 & 2.9509 & 2.6322 & 2.4480 & 3.9714 \\
   { 30$\leq$DTM$<$60} & 2.3502 & 3.1034 & 2.9061 & 2.9161 & 3.7961 \\
   { 60$\leq$DTM$<$90} & 2.4637 & 3.0227 & 3.3932 & 3.5187 & 3.5975 \\
   {90$\leq$DTM$<$120} & 2.6948 & 2.9482 & 3.7651 & 3.8349 & 3.6696 \\
   {120$\leq$DTM$<$150} & 2.4959 & 2.8202 & 4.2091 & 4.2501 & 3.6615 \\
   { 150$\leq$DTM} & 2.7251 & 2.9157 & 4.3636 & 4.3744 & 3.8217 \\
          &       &       &       &         &  \\
    \multicolumn{6}{l}{\it Partitioned by the level of VIX} \\[3pt]
   { VIX$<$15    } & 1.6727 & 2.0883 & 2.5067 & 2.4860 & 3.4768 \\
   { 15$\leq$VIX$<$20    } & 2.5730 & 3.1371 & 3.1802 & 2.7744 & 3.5961 \\
   { 20$\leq$VIX$<$25    } & 3.2861 & 3.9875 & 3.8534 & 3.5819 & 3.6284 \\
   { 25$\leq$VIX$<$30    } & 3.3513 & 4.3955 & 4.8263 & 4.7979 & 3.8781 \\
   { 30$\leq$VIX$<$35   } & 3.2173 & 4.6628 & 5.3971 & 5.9139 & 4.7166 \\
   { 35$\leq$VIX    } & 5.5187 & 5.9686 & 7.7183 & 9.6217 & 8.5992 \\
\\[0.0cm]
\\[-0.5cm]
\midrule
\bottomrule
\end{tabularx}
\end{footnotesize}
\par\end{centering}
{\small{}Note: This table reports the out-of-sample option pricing
performance. We evaluate the model's option pricing ability through
the root of mean square errors of implied volatility (${\rm RMSE_{IV}}$).
We summarize the results by option moneyness, maturity and market
VIX level. Moneyness is measured by Delta computed from the Black-Scholes
model. DTM denotes the number of calendar days to maturity. We conduct
our out-of-sample performance evaluation by equally splitting our
original dataset into two subsamples: the in-sample data consists
of the first 15 years (1990-2004) and the out-of-sample consists of
the remaining 15 years (2005-2019). The estimation for each model
is done only once for the in-sample data, and then price the out-of-sample
data by the estimated parameters.\label{tab:OOSOptionPricing}}{\small\par}
\end{table}

\subsubsection{Out-of-sample Option Pricing}

In this section, we turn to out-of-sample comparisons of the 
pricing models. A Markov switching model is more heavily parameterized
and this can lead to overfitting when the model is estimated and evaluated
with the same data. Here we follow \citet{ChristoffersenJacobs2004ms}
and split the original sample into two subsamples. We use the first
15 years, 1990-2004, exclusively for model estimation (in-sample)
and use the remaining 15 years, 2005-2019, for model evaluation (out-of-sample)
of the estimated models. So, each of the models is estimated exactly
once using the in-sample data and their abilities to price options
are exclusively evaluated over the out-of-sample period.

We report the out-of-sample RMSEs in Table \ref{tab:OOSOptionPricing}.
The relative ranking of the models is identical to the relative ranking
we found in-sample. The MS-RG model continues to be the best-performing
model. Not only is the MS-RG model the best on average, but it is
also the best within all subcategories. Impressively, the MS-RG has
nearly the same RMSE out-of-sample as it does in-sample. Given the
great out-of-sample performance by the MS-RG, we conclude that its
better option pricing performance is not a result of overfitting,
but rather reflects a real and substantive improvement in model-based
option pricing, which can be attributed to the time-varying risk aversion
that the MS-RG model can capture.

\section{Conclusion}

We introduced Markov-switching to the Realized GARCH model and combined
it with an exponentially affine pricing kernel with a time-varying aversion to volatility
risk. A key feature of this framework is that time-variation
in the pricing kernel is introduced with the same hidden Markov process that 
is used in the Realized GARCH model. In this way, the hidden
Markov chain process brings time-variation in both the physical
measure and the risk-neutral measure. We
derived model-implied pricing formula for European options in this framework. This was achieved with an analytical approximation method that is based on an Edgeworth
expansion of the density for cumulative return. 

The MS-RG model is straightforward to estimate from returns and realized
volatilities by quasi maximum likelihood estimation. Volatility models
with a hidden Markov-switching are typically challenging to estimate.
We circumvent the usual complications by introducing Markov switching
in a manner where states probabilities can be inferred from the
realized measures and returns. Estimating the parameters in the pricing kernel
is more involved and requires that option prices are included in the
empirical analysis. We estimated the model with a large panel of option
prices using 30 years of data (from 1990 to 2019)
and find that investors have a state-dependent tolerance for volatility-specific risk. Consistent with the existing literature, we find that
the volatility risk premium is, on average, negative. However, there
are periods when investors appear to have an appetite for variance
risk. These periods tend to coincide with relatively low levels of
the VIX. During other periods, we find that investors have a relatively
high aversion to volatility risk, and these periods are often preceded
by sudden upwards jumps in the VIX.

We compared the proposed model with several benchmarks and found it
to outperform all competing models. The option pricing model that is based
on the MS-RG framework leads to substantial improvements in option pricing. 
This reduction in the RMSE of option pricing errors range from 15.3\%
to 30.8\% relative to the alternative models. The same magnitudes of improvements
are seen across options with different characteristics. These improvements
are also seen out-of-sample, where the RMSE reductions
in option pricing errors range from 17\% to 34.5\%.

\section{Data Availability Statement}

The S\&P 500 index was collected from Yahoo Finance and CBOE VIX is from the CBOE website. The SPX options are available from the Optsum data (1990--1995) and OptionMetrics
(1996--2019). The realized measure of volatility was obtained from the Realized Library of Oxford-Man institute (2000-2019) and TickData (1990-1999).


\newpage{}

\appendix

\section{Appendix of Proofs\label{app:Appendix-of-Proofs}}

\setcounter{equation}{0}\renewcommand{\theequation}{A.\arabic{equation}}

\subsubsection*{Proof that $\psi=-\lambda$}

Imposing the no-arbitrage condition $\mathbb{E}_{t}^{\mathbb{Q}}[\exp(R_{t+1})]=\exp(r)$
yields 
\begin{eqnarray}
\mathbb{E}_{t}^{\mathbb{Q}}[\exp(R_{t+1})] & = & \mathbb{E}_{t}^{\mathbb{P}}[\mathbb{E}_{t}^{\mathbb{P}}[M_{t+1,t}(s_{t+1})\exp(R_{t+1})|s_{t+1}]]\label{eq:psi=00003Dminuslambda}\\
 & = & \exp\left[r+(\lambda+\psi)\mathbb{E}_{t}^{\mathbb{P}}\sqrt{h_{t+1}}\right]=\exp(r),\nonumber 
\end{eqnarray}
which shows that $\psi=-\lambda$.

\hfill{}$\square$

\noindent\textbf{Proof of Theorem \ref{thm:Qdynamics}}. The MGF
of $z_{t+1}^{\ast}\equiv z_{t+1}-\psi$ and $u_{t+1}^{\ast}=u_{t+1}-\chi_{s_{t+1}}$
under $\mathbb{Q}$ is 
\begin{eqnarray*}
 &  & \mathbb{E}_{t}^{\mathbb{Q}}\left[\exp(\theta_{1}z_{t+1}^{*}+\theta_{2}u_{t+1}^{*})\right]\\
 & = & \mathbb{E}_{t}^{\mathbb{P}}\left[M_{t+1,t}(s_{t+1})\exp(\theta_{1}z_{t+1}^{*}+\theta_{2}u_{t+1}^{*})\right]\\
 & = & \mathbb{E}_{t}^{\mathbb{P}}\left\{ \mathbb{E}_{t}^{\mathbb{P}}\left[M_{t+1,t}(s_{t+1})\exp(\theta_{1}z_{t+1}^{*}+\theta_{2}u_{t+1}^{*})|s_{t+1}\right]\right\} \\
 & = & \mathbb{E}_{t}^{\mathbb{P}}\left\{ \mathbb{E}_{t}^{\mathbb{P}}\left[M_{t+1,t}(s_{t+1})\exp(\theta_{1}z_{t+1}-\theta_{1}\psi+\theta_{2}u_{t+1}-\theta_{2}\chi_{s_{t+1}})|s_{t+1}\right]\right\} \\
 & = & \mathbb{E}_{t}^{\mathbb{P}}\left\{ \exp\left(\frac{1}{2}\left(\theta_{1}+\psi\right)^{2}+\frac{1}{2}\left(\theta_{2}+\chi_{s_{t+1}}\right)^{2}-\theta_{1}\psi-\theta_{2}\chi_{s_{t+1}}-\frac{1}{2}\psi^{2}-\frac{1}{2}\chi_{s_{t+1}}^{2}\right)\right\} \\
 & = & \exp\left(\frac{1}{2}\theta_{1}^{2}+\frac{1}{2}\theta_{2}^{2}\right)
\end{eqnarray*}
which shows $z_{t+1}^{\ast}\equiv z_{t+1}-\psi$ and $u_{t+1}^{\ast}=u_{t+1}-\chi_{s_{t+1}}$
are i.i.d. $N(0,1)$ in $\mathbb{Q}$.

\hfill{}$\square$

\noindent\textbf{Proof of Corollary \ref{cor:E(h_=00007Bt+n=00007D|s_t)}}.
Note that the dynamic of $\log h_{t}$ in $\mathbb{Q}$-measure can
be written as
\begin{align*}
\log h_{t+1} & =\underbrace{\omega^{*}+\gamma\xi_{s_{t}}^{*}}_{\zeta^{\prime}s_{t}}+\underbrace{(\beta+\gamma\phi)}_{\rho}\log h_{t}+\underbrace{(\tau_{1}^{*}+\gamma\delta_{1}^{*})z_{t}^{*}+(\tau_{2}+\gamma\delta_{2})(z_{t}^{*2}-1)+\gamma\sigma u_{t}^{*}}_{v_{t}}\\
 & =\zeta^{\prime}s_{t}+\rho\log h_{t}+v_{t}
\end{align*}
where $\zeta\in \mathbb{R}^{N\times 1}$, with $\zeta_j = \omega^{*}+\gamma\xi^{*}_j,\ j=1,\ldots,N $. The logarithm of the MGF of $v_{t}$, denoted as $G(\theta)=\log\mathbb{E}_{t}^{\mathbb{Q}}\left[\exp(\theta v_{t+1})\right]$,
is given by
\[
G(\theta)=-\frac{1}{2}\log(1-2\theta(\tau_{2}+\gamma\delta_{2}))-\theta(\tau_{2}+\gamma\delta_{2})+\frac{\theta^{2}(\tau_{1}^{*}+\gamma\delta_{1}^{*})^{2}}{2-4\theta(\tau_{2}+\gamma\delta_{2})}+\frac{1}{2}\gamma^{2}\sigma^{2}\theta^{2}.
\]
Note that the MGF of $s_{t}$, for any vector $\varphi \in \mathbb{R}^{N\times 1}$, is given by
\begin{eqnarray} 
\mathbb{E}_{t}^{\mathbb{Q}}\left[\left. \exp(\varphi^{\prime}s_{t+1})\right|s_{t}\right] & = & 
\mathbb{E}_{t}^{\mathbb{P}}\left[\left. \exp(\varphi^{\prime}s_{t+1})\right|s_{t}\right]  =  \mathbb{E}_{t}^{\mathbb{P}}\left[ \left. \exp\left(\sum_{i=1}^{N}\varphi_{i}s_{t+1,i}\right)\right| s_{t}\right]   \label{eq: MGFst} \\
&=&\sum_{i=1}^{N}\left(\sum_{j=1}^{N}\pi_{ij}\exp(\varphi_{j})\right)s_{t,i} \nonumber  =  \exp\left(\Delta^{\prime}(\varphi)s_{t}\right)
\end{eqnarray}
where the vector function $\Delta(\varphi)$ is defined by 
$$\Delta_{i}(\varphi)\equiv\log\left(\sum_{j=1}^{N}\pi_{ij}\exp(\varphi_{j})\right), \quad i=1,\ldots,N .$$ 

Next, we will derive the formula for a more general form given by 
$\mathbb{E}_{t}^{\mathbb{Q}}\left[h_{t+n}^{m}\exp(\varphi^{\prime}s_{t+n-1})|s_{t}\right]$. Note that for $n\geq2$, we have 
\begin{eqnarray*}
h_{t+n}^{m}=\exp\left(\sum_{k=1}^{n-1}m\rho^{n-1-k}\zeta^{\prime}s_{t+k}+m\rho^{n-1}\log h_{t+1}+\sum_{k=1}^{n-1}m\rho^{n-1-k}v_{t+k}\right)
\end{eqnarray*}
then 
\begin{align*}
\mathbb{E}_{t}\left[h_{t+n}^{m}\exp(\varphi^{\prime}s_{t+n-1})|s_{t}\right] & =\mathbb{E}_{t}\left[\left. \exp\left(\sum_{k=1}^{n-1}m\rho^{n-1-k}\zeta^{\prime}s_{t+k}+\varphi^{\prime}s_{t+n-1}+\sum_{k=1}^{n-1}G(m\rho^{n-1-k})\right)\right|s_{t} \right]h_{t+1}^{m\rho^{n-1}}\\
 & =\mathbb{E}_{t}^{\mathbb{Q}}\left[\left.\exp\left(\sum_{k=1}^{n-1}\phi_{(m,n-1-k,\varphi)}^{\prime}s_{t+k}\right)\right|s_{t}\right] h_{t+1}^{m\rho^{n-1}}
\end{align*}
where the vector $\phi_{(m,i,\varphi)}\in \mathbb{R}^{N\times 1}$ is defined as
\[
\phi_{(m,i,\varphi)}=\begin{cases}
G(m\rho^{i})+ m\rho^{i}\zeta, & \ i> 0\\
G(m)+ m\zeta+\varphi, & \ i=0.
\end{cases}
\]
Suppose that, for $n\geq2$, we have
\begin{align*}
\mathbb{E}_{t}^{\mathbb{Q}}\left[\left.\exp\left(\sum_{k=1}^{n-1}\phi_{(m,n-1-k,\varphi)}^{\prime}s_{t+k}\right)\right|s_{t}\right] & =\exp\left(\theta_{n}^{\prime}(m,\varphi)s_{t}\right),
\end{align*}
for $n=2$, we have $\theta_{2}(m,\varphi)=\Delta(G(m)+m\zeta+\varphi)$
from (\ref{eq: MGFst}). For $n=n+1$, we have
\begin{eqnarray*}
 &  & \mathbb{E}_{t}^{\mathbb{Q}}\left[\left.\exp\left(\sum_{k=1}^{n}\phi_{(m,n-k,\varphi)}^{\prime}s_{t+k}\right)\right|s_{t}\right]=\mathbb{E}_{t}^{\mathbb{Q}}\left[\left.\mathbb{E}_{t+1}^{\mathbb{Q}}\left(\exp(\sum_{k=1}^{n}\phi_{(m,n-k,\varphi)}^{\prime}s_{t+k})|s_{t+1}\right)\right|s_{t}\right]\\
 & = & \mathbb{E}_{t}^{\mathbb{Q}}\left[\left.\exp\left(\phi_{(m,n-1,\varphi)}^{\prime}s_{t+1}\right)\mathbb{E}_{t+1}^{\mathbb{Q}}\left(\exp(\sum_{k=2}^{n}\phi_{(m,n-k,\varphi)}^{\prime}s_{t+k})|s_{t+1}\right)\right|s_{t}\right]\\
 & = & \mathbb{E}_{t}^{\mathbb{Q}}\left[\left.\exp\left(\phi_{(m,n-1,\varphi)}^{\prime}s_{t+1}\right)\mathbb{E}_{t+1}^{\mathbb{Q}}\left(\exp(\sum_{k^{*}=1}^{n-1}\phi_{(m,n-1-k^{*},\varphi)}^{\prime}s_{t+1+k^{*}})|s_{t+1}\right)\right|s_{t}\right]\\
 & = & \mathbb{E}_{t}^{\mathbb{Q}}\left[\left.\exp\left(\phi_{(m,n-1,\varphi)}^{\prime}s_{t+1}+\theta_{n}^{\prime}(m,\varphi)s_{t+1}\right)\right|s_{t}\right]  =  \exp\left(\theta_{n+1}^{\prime}(m,\varphi)s_{t}\right),
\end{eqnarray*}
with following iterative process
\[
\theta_{n+1}(m,\varphi)=\Delta\left(\phi_{(m,n-1,\varphi)}+\theta_{n}(m,\varphi)\right).
\]
Now we obtain
\begin{align*}
\bm{\Psi}_{n}\left(m,\varphi\right)\equiv\text{\ensuremath{\mathbb{E}}}_{t}^{\mathbb{Q}}\left(h_{t+n}^{m}\exp(\varphi^{\prime}s_{t+n-1})|s_{t}\right) & =h_{t+1}^{m\rho^{n-1}}\exp\left(\theta_{n}^{\prime}(m,\varphi)s_{t}\right).
\end{align*}
Finally, the formula of $\mathbb{E}_{t}^{\mathbb{Q}}\left(h_{t+n}|s_{t}\right)$
can be obtained by setting $m=1,\varphi=\bm{0}$, such that
\[
\mathbb{E}_{t}^{\mathbb{Q}}\left(h_{t+n}|s_{t}\right)=\exp\left(\rho^{n-1}\log h_{t+1}+\theta_{n}^{\prime}s_{t}\right) = \exp\left(\kappa_n + \rho^{n-1}\log h_{t+1}+\tilde{\theta}_{n}^{\prime}s_{t}\right),
\]
with $\kappa_n = \sum_{i=0}^{n-2}G(\rho^{i})$, $\tilde{\theta}_{1}=\bm{0}$ and $\tilde{\theta}_{n+1}=\Delta\left(\rho^{n-1}\zeta+\tilde{\theta}_{n}\right)$.\hfill{}$\square$

\section{Terms for Analytical Approximation\label{sec:Terms-for-Analytical}}

\setcounter{equation}{0}\renewcommand{\theequation}{B.\arabic{equation}}

To simplify the expression, we use the notation $\mathbb{E}_{t}(\cdot)=\mathbb{E}_{t}^{\mathbb{Q}}(\cdot|s_{t})$
below. Without loss of generality, we set $t=0$, such that $T$ is
the number of days to maturity. Now, the conditional moment of cumulative returns $R_T \equiv \log(S_T/S_0)$ is expressed as
\[
\ensuremath{\mathbb{E}_{0}\left(R_{T}^{j}\right)=\mathbb{E}_{0}\left[\sum_{i=1}^{T}\left(r-\frac{1}{2}h_{t+i}+\sqrt{h_{t+i}}z_{t+i}\right)^{j}\right].}
\]
Expanding the formula, we get
\begin{eqnarray*}
\mathbb{E}_{0}\left(R_{T}\right) & = & Tr-\frac{1}{2}\sum_{i=1}^{T}\mathbb{E}_{0}\left[h_{i}\right],\\
\mathbb{E}_{0}\left(R_{T}^{2}\right) & = & T^{2}r^{2}-Tr\sum_{i=1}^{T}\mathbb{E}_{0}\left[h_{i}\right]+\frac{1}{4}S_{D_{1}}+S_{D_{2}}-S_{D_{3}},\\
\mathbb{E}_{0}\left(R_{T}^{3}\right) & = & \ensuremath{T^{3}r^{3}-\frac{3}{2}T^{2}r^{2}\sum_{i=1}^{T}\mathbb{E}_{0}\left[h_{i}\right]+3Tr\left(\frac{1}{4}S_{D_{1}}+S_{D_{2}}-S_{D_{3}}\right)}\\
 &  & +\left(-\frac{1}{8}S_{T_{1}}+S_{T_{2}}+\frac{3}{4}S_{T_{3}}-\frac{3}{2}S_{T_{4}}\right),\\
\mathbb{E}_{0}\left(R_{T}^{4}\right) & = & T^{4}r^{4}-2T^{3}r^{3}\sum_{i=1}^{T}\mathbb{E}_{0}\left[h_{i}\right]+6T^{2}r^{2}\left(\frac{1}{4}S_{D_{1}}+S_{D_{2}}-S_{D_{3}}\right)\\
 &  & +Tr\left(-\frac{1}{2}S_{T_{1}}+4S_{T_{2}}+3S_{T_{3}}-6S_{T_{4}}\right) +\left(\frac{1}{16}S_{Q_{1}}+S_{Q_{2}}-\frac{1}{2}S_{Q_{3}}+\frac{3}{2}S_{Q_{4}}-2S_{Q_{5}}\right) 
\end{eqnarray*}
The formula of $S_{D_{i}}$, $S_{T_{i}}$ and $S_{Q_{i}}$ are summations
of some expectations related to future volatility and shocks that
were derived in \citet[p. 104]{DuanGauthierSimonato1999}.
For instance, the formula of $S_{T_{2}}$ is given by
\[
\ensuremath{S_{T_{2}}=\mathbb{E}_{0}\left[\sum_{i=1}^{T}\sum_{j=1}^{T}\sum_{k=i}^{T}\sqrt{h_{i}}z_{i}\sqrt{h_{j}}z_{j}\sqrt{h_{k}}z_{k}\right]=3\sum_{i=1}^{T}\sum_{j=1}^{T-i}\mathbb{E}_{0}\left[\sqrt{h_{i}}z_{i}h_{i+j}\right]}.
\]
The expression requires 19 types of terms as input, which must be
derived for the present model. For instance, one type of these terms take the
form, $\mathbb{E}_{0}\left[\sqrt{h_{i}}z_{i}h_{i+j}\right]$, and
it is needed for the computation of $S_{T_{2}}$ above. Each of the
19 terms are derived below, and numbered with (\ref{eq:B1})-(\ref{eq:B17}). 

\subsection*{Expectations without $z$}

We will first derive the formula for following function
\begin{eqnarray*}
\mathbb{E}_{0}\left(h_{i}^{k}h_{i+j}^{m}\right) & = & \mathbb{E}_{0}\left[h_{i}^{k}\mathbb{E}_{i-1}(h_{i+j}^{m})\right]\\
 & = & \mathbb{E}_{0}\left[h_{i}^{k}h_{i}^{m\rho^{j}}e^{\theta_{j+1}^{\prime}(m,0)s_{i-1}}\right]\\
 & = & h_{1}^{(k+m\rho^{j})\rho^{i-1}}e^{\theta_{i}^{\prime}(k+m\rho^{j},\theta_{j+1}(m,0))s_{0}}\\
 & = & \bm{\Psi}_{i}\left(k+m\rho^{j},\theta_{j+1}(m,0)\right).
\end{eqnarray*}
Following \citet{DuanGauthierSimonato1999}, the terms needed for
analytical approximation include:
\begin{eqnarray}
\mathbb{E}_{0}\left(h_{i}^m\right) & = & \bm{\Psi}_{i}\left(m,0\right),\label{eq:B1}\\
\mathbb{E}_{0}\left(h_{i}h_{i+j}\right) & = & \bm{\Psi}_{i}\left(1+\rho^{j},\theta_{j+1}(1,0)\right), \\
\mathbb{E}_{0}\left(h_{i}^{2}h_{i+j}\right) & = & \bm{\Psi}_{i}\left(2+\rho^{j},\theta_{j+1}(1,0)\right),\\
\mathbb{E}_{0}\left(h_{i}h_{i+j}^{2}\right) & = & \bm{\Psi}_{i}\left(1+2\rho^{j},\theta_{j+1}(2,0)\right),
\end{eqnarray}
and
\begin{eqnarray}
\mathbb{E}_{0}\left(h_{i}h_{i+j}h_{i+j+k}\right) & = & \mathbb{E}_{0}\left[h_{i}\mathbb{E}_{i-1}\left(h_{i+j}h_{i+j+k}\right)\right]\nonumber \\
 & = & \mathbb{E}_{0}\left[h_{i}h_{i}^{(1+\rho^{k})\rho^{j}}e^{\theta_{j+1}^{\prime}(1+\rho^{k},\theta_{k+1}(1,0))s_{i-1}}\right]\nonumber \\
 & = & \bm{\Psi}_{i}\left(1+\rho^{j}+\rho^{k+j},\theta_{j+1}(1+\rho^{k},\theta_{k+1}(1,0))\right).
\end{eqnarray}

\subsection*{Expectations with $z$}

For terms were $z$ is involved, we define the function
\begin{eqnarray*}
\bm{\Gamma}_{i}\left(j,r,m,\varphi\right) & \equiv & \mathbb{E}_{0}\left(h_{i}^{\tfrac{r}{2}}z_{i}^{r}h_{i+j}^{m}e^{\varphi^{\prime}s_{i+j-1}}\right)\\
 & = & \mathbb{E}_{0}\left(h_{i}^{\tfrac{r}{2}}z_{i}^{r}\mathbb{E}_{i}\left(h_{i+j}^{m}e^{\varphi^{\prime}s_{i+j-1}}\right)\right)\\
 & = & \mathbb{E}_{0}\left(h_{i}^{\tfrac{r}{2}}z_{i}^{r}h_{i+1}^{m\rho^{j-1}}e^{\theta_{j}^{\prime}(m,\varphi)s_{i}}\right)\\
 & = & \mathbb{E}_{0}\left(h_{i}^{\tfrac{r}{2}}z_{i}^{r}e^{\left(m\rho^{j-1}\zeta+\theta_{j}(m,\varphi)\right)^{\prime}s_{i}+m\rho^{j-1}\rho\log h_{i}+m\rho^{j-1}v_{i}}\right)\\
 & = & \mathbb{E}_{0}\left[h_{i}^{\tfrac{r}{2}+m\rho^{j}}e^{\left(m\rho^{j-1}\zeta+\theta_{j}(m,\varphi)\right)^{\prime}s_{i}}\right]\mathbb{E}_{0}\left(z_{i}^{r}e^{m\rho^{j-1}v_{i}}\right)\\
 & = & \mathbb{E}_{0}\left[h_{i}^{\tfrac{r}{2}+m\rho^{j}}e^{\Delta^{\prime}\left(m\rho^{j-1}\zeta+\theta_{j}(m,\varphi)\right)s_{i-1}}\right]\mathbb{E}_{0}\left(z_{i}^{r}e^{m\rho^{j-1}v_{i}}\right)\\
 & = & h_{1}^{a_{i}(j,r,m)}e^{b_{i}^{\prime}(j,r,m,\varphi)s_{0}}c_{i}(j,r,m),
\end{eqnarray*}
where
\begin{eqnarray*}
a_{i}(j,r,m) & = & \left(\tfrac{r}{2}+m\rho^{j}\right)\rho^{i-1},\\
b_{i}(j,r,m,\varphi) & = & \theta_{i}\left(\tfrac{r}{2}+m\rho^{j},\Delta\left(m\rho^{j-1}\zeta+\theta_{j}(m,\varphi)\right)\right),\\
c_{i}(j,r,m) & = & \mathbb{E}_{0}\left(z_{i}^{r}e^{m\rho^{j-1}v_{i}}\right).
\end{eqnarray*}
The expression of $\mathbb{E}_{0}\left(z_{i}^{r}e^{kv_{i}}\right)$
is provided in \citet[p. 354]{HuangWangHansen2017}. So, we have:
\begin{eqnarray}
\mathbb{E}_{0}\left(\sqrt{h_{i}}z_{i}h_{i+j}\right) & = & \bm{\Gamma}_{i}\left(j,1,1,0\right),\\
\mathbb{E}_{0}\left(\sqrt{h_{i}}z_{i}h_{i+j}^{2}\right) & = & \bm{\Gamma}_{i}\left(j,1,2,0\right),\\
\mathbb{E}_{0}\left(h_{i}z_{i}^{2}h_{i+j}\right) & = & \bm{\Gamma}_{i}\left(j,2,1,0\right),\\
\mathbb{E}_{0}\left(h_{i}^{3/2}z_{i}^{3}h_{i+j}\right) & = & \bm{\Gamma}_{i}\left(j,3,1,0\right),
\end{eqnarray}
\begin{eqnarray}
\ensuremath{\mathbb{E}_{0}\left(h_{i}\sqrt{h_{i+j}}z_{i+j}h_{i+j+k}\right)} & = & \mathbb{E}_{0}\left(h_{i}\mathbb{E}_{i-1}\left(\sqrt{h_{i+j}}z_{i+j}h_{i+j+k}\right)\right)\nonumber \\
 & = & \mathbb{E}_{0}\left(h_{i}^{1+a_{j+1}(k,1,1)}e^{b_{j+1}^{\prime}\left(k,1,1,0\right)s_{i-1}}c_{j+1}(k,1,1)\right)\nonumber \\
 & = & \bm{\Psi}_{i}\left(1+a_{j+1}\left(k,1,1\right),b_{j+1}\left(k,1,1,0\right)\right)c_{j+1}(k,1,1),
\end{eqnarray}
\begin{eqnarray}
\ensuremath{\mathbb{E}_{0}\left(\sqrt{h_{i}}z_{i}h_{i+j}h_{i+j+k}\right)} & = & \ensuremath{\mathbb{E}_{0}\left(\sqrt{h_{i}}z_{i}h_{i+j}\mathbb{E}_{i+j-1}(h_{i+j+k})\right)}\nonumber \\
 & = & \mathbb{E}_{0}\left(\sqrt{h_{i}}z_{i}h_{i+j}^{1+\rho^{k}}e^{\theta_{k+1}^{\prime}(1,0)s_{i+j-1}}\right)\nonumber \\
 & = & \bm{\Gamma}_{i}\left(j,1,1+\rho^{k},\theta_{k+1}(1,0)\right),
\end{eqnarray}
\begin{eqnarray}
\mathbb{E}_{0}\left(h_{i}^{3/2}z_{i}h_{i+j}\right) & = & \mathbb{E}_{0}\left(h_{i}\mathbb{E}_{i-1}\left(\sqrt{h_{i}}z_{i}h_{i+j}\right)\right)\nonumber \\
 & = & \mathbb{E}_{0}\left(h_{i}^{1+a_{1}(j,1,1)}e^{b_{1}^{\prime}(j,1,1,0)s_{i-1}}c_{1}(j,1,1)\right)\nonumber \\
 & = & \bm{\Psi}_{i}\left(1+a_{1}(j,1,1),b_{1}(j,1,1,0)\right)c_{1}(j,1,1),
\end{eqnarray}
\begin{eqnarray}
{\mathbb{E}_{0}\left(\sqrt{h_{i}}z_{i}\sqrt{h_{i+j}}z_{i+j}h_{i+j+k}\right)} & = & \mathbb{E}_{0}\left(\sqrt{h_{i}}z_{i}\mathbb{E}_{i}\left(\sqrt{h_{i+j}}z_{i+j}h_{i+j+k}\right)\right) \nonumber \\
 & = & \mathbb{E}_{0}\left(\sqrt{h_{i}}z_{i}h_{i+1}^{a_{j}(k,1,1)}e^{b_{j}^{\prime}(k,1,1,0)s_{i}}\right)c_{j}(k,1,1) \nonumber \\
 & = & \bm{\Gamma}_{i}\left(1,1,a_{j}\left(k,1,1\right),b_{j}\left(k,1,1,0\right)\right)c_{j}(k,1,1),
\end{eqnarray}
\begin{eqnarray}
\ensuremath{} &  & \mathbb{E}_{0}\left(\sqrt{h_{i}}z_{i}\sqrt{h_{i+j}}z_{i+j}h_{i+j+k}h_{i+j+k+m}\right)\nonumber \\
 & = & \mathbb{E}_{0}\left(\sqrt{h_{i}}z_{i}\mathbb{E}_{i}\left(\sqrt{h_{i+j}}z_{i+j}h_{i+j+k}h_{i+j+k+m}\right)\right)\nonumber \\
 & = & \mathbb{E}_{0}\left(\sqrt{h_{i}}z_{i}h_{i+1}^{a_{j}(k,1,1+\rho^{m})}e^{b_{j}^{\prime}(k,1,1+\rho^{m},\theta_{m+1}(1,0))s_{i}}\right)c_{j}(1,1,1+\rho^{m})\nonumber \\
 & = & \bm{\Gamma}_{i}\left(1,1,a_{j}(k,1,1+\rho^{m}),b_{j}(k,1,1+\rho^{m},\theta_{m+1}(1,0))\right)c_{j}(1,1,1+\rho^{m}),
\end{eqnarray}
\begin{eqnarray}
\mathbb{E}_{0}\left(h_{i}z_{i}^{2}h_{i+j}h_{i+j+k}\right) & = & \bm{\Gamma}_{i}\left(j,2,1+\rho^{k},\theta_{k+1}(1,0)\right),\\
\ensuremath{\mathbb{E}_{0}\left(h_{i}h_{i+j}z_{i+j}^{2}h_{i+j+k}\right)} & = & \bm{\Psi}_{i}\left(1+a_{j+1}\left(k,2,1\right),b_{j+1}\left(k,2,1,0\right)\right)c_{j+1}\left(k,2,1,0\right), \ \ \\
\ensuremath{\mathbb{E}_{0}\left(\sqrt{h_{i}}z_{i}h_{i+j}z_{i+j}^{2}h_{i+j+k}\right)} & = & \bm{\Gamma}_{i}\left(1,1,a_{j}(k,2,1),b_{j}(k,2,1,0)\right)c_{j}(k,2,1),
\end{eqnarray}
\begin{eqnarray}
\ensuremath{} &  & \mathbb{E}_{0}\left(\sqrt{h_{i}}z_{i}\sqrt{h_{i+j}}z_{i+j}\sqrt{h_{i+j+k}}z_{i+j+k}h_{i+j+k+m}\right)\nonumber \\
 & = & \mathbb{E}_{0}\left(\sqrt{h_{i}}z_{i}\mathbb{E}_{i}\left(\sqrt{h_{i+j}}z_{i+j}\sqrt{h_{i+j+k}}z_{i+j+k}h_{i+j+k+m}\right)\right)\nonumber \\
 & = & \mathbb{E}_{0}\left(\sqrt{h_{i}}z_{i}h_{i+1}^{a_{j}(1,1,a_{k}(m,1,1))}e^{b_{j}^{\prime}(1,1,a_{k}(m,1,1),b_{k}(m,1,1,0))s_{i}}\right)\nonumber \\
 &  & \quad\times c_{j}(1,1,a_{k}(m,1,1))c_{k}(m,1,1)\nonumber \\
 & = & \bm{\Gamma}_{i}\left(1,1,a_{j}(1,1,a_{k}(m,1,1)),b_{j}(1,1,a_{k}(m,1,1),b_{k}(m,1,1,0))\right)\nonumber \\
 &  & \quad\times c_{j}(1,1,a_{k}(m,1,1))c_{k}(m,1,1),
\end{eqnarray}
and finally,
\begin{eqnarray}
\ensuremath{\mathbb{E}_{0}\left(h_{i}z_{i}^{2}\sqrt{h_{i+j}}z_{i+j}h_{i+j+k}\right)} & = & \bm{\Gamma}_{i}\left(1,2,a_{j}(k,1,1),b_{j}(k,1,1,0)\right)c_{j}(k,1,1).\label{eq:B17}
\end{eqnarray}
\end{document}